\documentclass[preprint,aps,floatfix,nofootinbib,a4paper]{revtex4-1}
\pdfoutput=1

\usepackage{amsmath, amssymb, amsfonts, amsthm, latexsym, epsfig, mathrsfs, xcolor, bbm, slashed}

\usepackage[inline]{enumitem}

\usepackage{setspace}
\usepackage[marginal, multiple]{footmisc}

\usepackage[T1]{fontenc}
\usepackage[utf8]{inputenc}
\usepackage{lmodern}

\usepackage[colorlinks, allcolors=blue!70!black, linktocpage]{hyperref}

\numberwithin{equation}{section}

\usepackage{cleveref}

\usepackage{microtype}

\usepackage{floatrow}
\let\OLDtableofcontents\tableofcontents
\renewcommand\tableofcontents[1]{%
    {\baselineskip 0.5ex %
	\OLDtableofcontents{#1}}%
}

\let\OLDthebibliography\thebibliography
\renewcommand\thebibliography[1]{%
	\setstretch{1.079} 
	\OLDthebibliography{#1}%
	\small %
	\setlength{\itemsep}{0.2\baselineskip} 
}

\let\OLDfootnote\footnote
\renewcommand\footnote[1]{%
	\setlength{\footnotesep}{0.75\baselineskip}%
	{\footnotesize \OLDfootnote{#1}}%
}

\setlist[enumerate]{noitemsep, label=(\arabic*), ref=(\arabic*)}

\newlist{condlist}{enumerate}{2}
\setlist[condlist,1]{noitemsep, topsep=0pt, label=(\arabic*), ref=(\arabic*)}
\setlist[condlist,2]{noitemsep, label=(\alph*), ref=(\arabic{condlisti}.\alph*)}
\crefname{condlisti}{condition}{conditions}
\crefname{condlistii}{condition}{conditions}

\newlist{propertylist}{enumerate}{1}
\setlist[propertylist,1]{noitemsep, topsep=0pt, label=(\arabic*), ref=(\arabic*)}
\crefname{propertylisti}{Property}{Properties}

\renewcommand\thesection{\arabic{section}}
\renewcommand\thesubsection{\arabic{subsection}}

\makeatletter
\def\p@subsection{\thesection.}
\def\p@subsubsection{\thesection.\thesubsection.}
\makeatother 

\theoremstyle{plain}

\theoremstyle{definition}
\newtheorem{definition}{Definition}[section]

\theoremstyle{remark}
\newtheorem{remark}{Remark}[section]

\crefname{equation}{Eq.}{Eqs.}
\creflabelformat{equation}{#2#1#3}

\crefname{section}{\S}{\S}
\crefname{appendix}{Appendix}{Appendices}
\crefname{figure}{Fig.}{Figs.}
\crefname{table}{Table}{Tables}

\crefname{definition}{Def.}{Defs.}
\crefname{prop}{Prop.}{Props.}
\crefname{lemma}{Lemma}{Lemmas}
\crefname{corollary}{Cor.}{Cors.}
\crefname{thm}{Theorem}{Theorems}
\crefname{remark}{Remark}{Remarks}

\crefname{ass}{Assumptions}{Assumptions}
\crefname{property}{Properties}{Properties}

\newcommand{\be}{\begin{equation}}
\newcommand{\ee}{\end{equation}}

\newcommand{\lb}{\left}
\newcommand{\rb}{\right}

\newcommand{\mc}{\mathcal}

\newcommand{\ms}{\mathscr}
\newcommand{\mf}{\mathfrak}
\newcommand{\bb}{\mathbb}

\newcommand{\set}[1]{\lb\{ #1 \rb\}}
\newcommand{\st}{~:~}

\newcommand{\eqsp}{\, ,\quad} 

\newcommand{\hr}{\begin{center}* * *\end{center}}

\newcommand{\Lie}{\pounds} 
\newcommand{\defn}{\mathrel{\mathop:}=} 

\newcommand{\union}{\cup} 
\newcommand{\inter}{\cap} 

\let\oldsetminus\setminus
\renewcommand{\setminus}{\!\oldsetminus\!} 

\let\oldint\int
\renewcommand{\int}{\oldint\limits}

\let\oldlim\lim
\renewcommand{\lim}{\oldlim\limits}

\renewcommand{\bar}{\overline}

\newcommand{\scri}{\ms I}
\newcommand{\hyp}{\ms H}

\newcommand{\cyl}{\ms C}
\newcommand{\nulls}{\ms N}

\newcommand{\dd}[1]{\boldsymbol{#1}} 

\newcommand{\df}[1]{#1} 

\newcommand{\nfrac}[2]{{{}^#1\!\!/\!_#2}}

\newcommand{\half}{\nfrac{1}{2}}

\newcommand{\Omh}{\Omega^\half} 

\newcommand{\rsub}[1]{\scriptscriptstyle{\rm #1}}

\newcommand{\bs}{{\rsub{(BS)}}}
\newcommand{\cd}{{\rsub{(CD)}}}

\DeclareMathOperator{\thorn}{\text{\rm \th}}

\renewcommand{\Re}{{\rm Re\,}}

\let\thorn\relax
\DeclareMathOperator{\thorn}{\text{\rm \th}}
\let\eth\relax
\DeclareMathOperator{\eth}{\text{\rm \dh}}

\newcommand{\wt}{\circeq}

\begin{document}

\setstretch{1.2}


\title{Conservation of asymptotic charges from past to future null infinity: Supermomentum in general relativity}
\author{Kartik Prabhu}\email{kartikprabhu@cornell.edu}
\affiliation{Cornell Laboratory for Accelerator-based Sciences and Education (CLASSE),\\ Cornell University, Ithaca, NY 14853, USA}

\begin{abstract}
We show that the BMS-supertranslations and their associated supermomenta on past null infinity can be related to those on future null infinity, proving the conjecture of Strominger for a class of spacetimes which are asymptotically-flat in the sense of Ashtekar and Hansen. Using a cylindrical \(3\)-manifold of both null and spatial directions of approach towards spatial infinity, we impose appropriate regularity conditions on the Weyl tensor near spatial infinity along null directions. The asymptotic Einstein equations on this \(3\)-manifold and the regularity conditions imply that the relevant Weyl tensor components on past null infinity are antipodally matched to those on future null infinity. The subalgebra of totally fluxless supertranslations near spatial infinity provides a natural isomorphism between the BMS-supertranslations on past and future null infinity. This proves that the flux of the supermomenta is conserved from past to future null infinity in a classical gravitational scattering process provided additional suitable conditions are satisfied at the timelike infinities.
\end{abstract}

\maketitle
\tableofcontents

\section{Introduction}
\label{sec:intro}

For asymptotically-flat spacetimes describing isolated systems in general relativity, it is well-known that at both future and past null infinities one obtains an infinite-dimensional asymptotic symmetry group --- the Bondi-Metzner-Sachs (BMS) group --- along with the corresponding charges and fluxes due to gravitational radiation \cite{BBM, Sachs1, Sachs2, Penrose, GW, AS-symp, WZ}. Similarly, at spatial infinity one again obtains an infinite-dimensional asymptotic symmetry group --- the Spi-group --- and the Arnowitt-Deser-Misner (ADM) energy and angular momentum as conserved charges corresponding to a Poincar\'e subgroup \cite{ADM, ADMG, CR, Beig-Schmidt, AH, Sommers, Ash-in-Held, Ash-Rom, Friedrich, AES}. For a detailed review of asymptotic structures in general relativity see \cite{Geroch-asymp}.

Recently it has been conjectured by Strominger \cite{Stro-CK-match} that the (a priori independent) BMS groups on past and future null infinities can be related through an antipodal reflection near spatial infinity. Such a matching gives a \emph{global} ``diagonal'' asymptotic symmetry group for general relativity. If similar matching conditions (which were assumed as ``boundary conditions'' in \cite{Stro-CK-match}) relate the gravitational fields, it would imply infinitely many conservation laws in classical gravitational scattering in the sense that the incoming fluxes associated to the BMS group at past null infinity would equal the outgoing fluxes of the corresponding BMS group at future null infinity. It has been further conjectured that this diagonal group is also a symmetry of the scattering matrix in quantum gravity \cite{Stro-CK-match} and the corresponding conservation laws (in linearised gravity around Minkowski spacetime) have been related to various soft theorems \cite{HLMS, SZ}. These conservation laws are also speculated to play a role in the resolution of the black hole information loss problem \cite{HPS, Hawking} (see however \cite{BP} for a contrarian view).

However, the validity of such matching conditions for the asymptotic symmetries and charges has not been proven even in classical general relativity except in certain special cases discussed below. The main difficulty in resolving the matching problem is the limited structure available at spatial infinity in general spacetimes. The asymptotic behaviour of the gravitational field for any asymptotically-flat spacetime can be conveniently described in a conformally-related unphysical spacetime, the Penrose conformal-completion. In the unphysical spacetime, null infinities \(\scri^\pm\) are smooth null boundaries while spatial infinity is a boundary point \(i^0\) acting as the vertex of ``the light cone at infinity''. For Minkowski spacetime the unphysical spacetime is smooth (in fact, analytic) at \(i^0\), and so a natural identification exists between the null generators (and fields) on \(\scri^-\) with those on \(\scri^+\) by ``passing through'' \(i^0\). However, in more general spacetimes, the unphysical metric is not even once-differentiable at spatial infinity unless the ADM mass of the spacetime vanishes \cite{AH}, and the unphysical spacetime manifold \emph{does not} have a smooth differential structure at \(i^0\). Thus the identification between the null generators of \(\scri^-\) and \(\scri^+\), and the corresponding symmetries and fields, becomes much more difficult.

Nevertheless, Ashtekar and Magnon-Ashtekar \cite{Ash-Mag-Ash} showed that the limit of Bondi energy-momentum to \(i^0\) along both \(\scri^\pm\) equals the ADM energy-momentum. Similar result for the Bondi and ADM angular momentum was obtained by Ashtekar and Streubel \cite{AS-ang-mom} for spacetimes which are stationary near \(i^0\). For general supertranslations, the antipodal matching of all the infinite number of symmetries and charges has been shown in linearised gravity around a Minkowski background spacetime by Troessaert \cite{Tro}.\footnote{The result of \cite{Tro} can be viewed as the linearisation around a Minkowski background of the more general analysis in \cite{HL-GR-matching}.} In full nonlinear general relativity, it was argued by Strominger \cite{Stro-CK-match} that a similar result should hold for all supertranslations in the Christodoulou-Klainerman class of spacetimes (CK-spacetimes) \cite{CK}. However, in such spacetimes only the Bondi energy-momentum associated to translations is non-vanishing in the limit to \(i^0\) along \(\scri\) and the supermomenta associated to the rest of the supertranslations vanish (see \cite{Ash-CK-triv}). Thus, although CK-spacetimes form an open ball in some suitable topology around Minkowski spacetime, they are not general enough to address the non-trivial aspects of the matching problem for supertranslations. A non-trivial result in the nonlinear theory was obtained by Herberthson and Ludvigsen \cite{HL-GR-matching} who proved that the Weyl tensor component entering the Bondi mass formula (denoted by \(\psi_2\) in \cref{sec:non-Bondi}) on \(\scri^-\) matches antipodally with the corresponding quantity on \(\scri^+\) on spacetimes which are sufficiently regular in the limit to \(i^0\) along \(\scri\). The matching of the supertranslation symmetries was not addressed in \cite{HL-GR-matching}, but assuming an antipodal matching of supertranslations (as proposed in \cite{Stro-CK-match}), the earlier result of \cite{HL-GR-matching} resolves the matching problem for a class of spacetimes (more general than the ones in \cite{Stro-CK-match}) where the News tensor and certain connection components falloff fast enough in the limit to \(i^0\). We note that a key improvement in \cite{HL-GR-matching,Tro} over \cite{Stro-CK-match} is that the antipodal matching of the relevant fields is not imposed a priori as a boundary condition, but follows from the regularity of solutions to the Einstein equation on \(\scri\) and at \(i^0\).\\

In this paper we prove the matching conditions for \emph{all} asymptotic supertranslations in general relativity on asymptotically-flat spacetimes satisfying suitably regularity conditions near spatial infinity (see \cref{def:null-regular}). We will use the methods of \cite{KP-EM-match} where an analogous result was shown for Maxwell fields on any asymptotically-flat spacetime. our result can be viewed as a generalisation of \cite{Ash-Mag-Ash} to include all supertranslations, or of \cite{Tro} to full nonlinear general relativity, or of \cite{HL-GR-matching} to show the antipodal matching of not only the Weyl tensor but also the relevant supertranslation symmetries. 

Both null and spatial infinities can be treated in a unified spacetime-covariant manner using the definition of asymptotic-flatness given by Ashtekar and Hansen \cite{AH, Ash-in-Held}  (\cref{def:AH}). In the Ashtekar-Hansen formalism, instead of working directly at the point \(i^0\) where sufficiently smooth structure is unavailable, one works on the space of spatial directions at \(i^0\) given by a timelike-unit-hyperboloid \(\hyp\) in the tangent space at \(i^0\) (\cref{fig:hyperboloid}). The Weyl tensor of the unphysical spacetime (suitably conformally rescaled) admits limits to \(i^0\) which depend on the direction of approach and thus induces \emph{smooth} fields on \(\hyp\). The asymptotic Spi-supertranslations at spatial infinity then give us infinitely many charges on \(\hyp\) in terms of these smooth limiting fields.

However, for the matching problem we are interested in the behaviour of fields at \(i^0\) along \(\scri^\pm\) i.e., along null directions at \(i^0\). Such null directions can be incorporated into the Ashtekar-Hansen formalism using a space of both null and spatial directions of approach to \(i^0\) constructed in \cite{KP-EM-match}. This space is a cylinder \(\cyl\) in the tangent space at \(i^0\), which is diffeomorphic to a conformal-completion of \(\hyp\) (see \cref{fig:cylinder}). The two boundaries \(\nulls^\pm\)  of \(\cyl\) correspond to the directions of approach to \(i^0\) in null directions along \(\scri^\pm\). Using this diffeomorphism, we can study the asymptotic gravitational fields and supertranslations on \(\cyl\), instead of on \(\hyp\). There is a reflection map, which acts as a conformal isometry on \(\cyl\) \cref{eq:reflection-cyl}, which allows us to identify the null generators of \(\scri^\pm\) represented by the spaces \(\nulls^\pm\) of null directions at \(i^0\).

With this geometric setup we can ask about two different limits of the gravitational fields and supertranslations:
\begin{enumerate*}
    \item first take the limit to \(\scri^\pm\) and then towards \(i^0\), or,
    \item first take the limit to \(i^0\) along spatial directions (now represented by \(\cyl\setminus\nulls^\pm\)) and then take the limit where the direction of approach becomes null i.e., a limit to \(\nulls^\pm\).
\end{enumerate*}
In general, neither of these limits might exist given the conditions by Ashtekar and Hansen. Thus, we impose additional \emph{null-regularity} conditions on some of the gravitational fields (\cref{def:null-regular}). These conditions imply that both limits, taken as described above, exist and the induced limiting fields on the boundaries \(\nulls^\pm\) obtained by both limiting procedures match. Thus, the null-regularity conditions act as ``continuity'' conditions on the gravitational fields at \(i^0\) and further ensure that the flux of charges on \(\scri^\pm\) is finite. This will lead us to a partial matching of the supertranslation symmetries, whereby any BMS-supertranslation on \(\scri^\pm\) gives some (not unique) Spi-supertranslation on \(\cyl\) such that they match continuously at \(\nulls^\pm\).

Using the asymptotic Einstein equations on \(\cyl\) we show that, with our null-regularity conditions, the fields entering the expression for the charges from the past null directions \(\nulls^-\) match antipodally to those from the future null directions \(\nulls^+\) (under the reflection map on \(\cyl\)). Finally, we isolate a subalgebra of Spi-supertranslations for which the total flux of the corresponding supermomenta across \emph{all} of \(\cyl\) vanishes. This corresponds to the physical requirement that in a scattering process one only is concerned with the fields on null infinity, and any flux through spatial infinity is ``non-dynamical''. We emphasise that this is not a restriction on the kinds of spacetimes we consider (unlike the previously discussed null-regularity conditions), but a choice of supertranslations relevant to a scattering process. Such totally fluxless Spi-supertranslations on \(\cyl\) then give us the desired isomorphism between the BMS-supertranslations and a conservation law for the fluxes of BMS-supermomenta between \(\scri^-\) and \(\scri^+\), proving the conjecture in \cite{Stro-CK-match}.

\hr

The rest of the paper is organised as follows. In \cref{sec:AH} we review the Ashtekar-Hansen structure of spacetimes that are asymptotically-flat at both null and spatial infinity. In \cref{sec:cyl}, we summarise the construction of the space \(\cyl\) at spatial infinity that includes both null and spatial directions and its relation to the unit-hyperboloid \(\hyp\) in the Ashtekar-Hansen framework. In \cref{sec:super-trans} we review the asymptotic supertranslation symmetries and the associated supermomenta at \(\scri\) and in the Ashtekar-Hansen formalism at \(i^0\). In \cref{sec:matching}, we impose suitable regularity conditions on the gravitational fields on null infinity which ensure that the BMS-supermomenta defined on null infinity remain finite as we approach spatial infinity. We introduce the subalgebra of Spi-supertranslations for which the total flux of Spi-supermomenta across spatial infinity vanishes, which then gives us the antipodal matching conditions, the global diagonal symmetry algebra and the flux conservation between past and future null infinity. We end with \cref{sec:disc} summarising and discussing our analysis and results.

 We collect the computations generalising the relevant field equations on null infinity to arbitrary conformal choices in \cref{sec:non-Bondi}. In \cref{sec:dd} we collect the definitions of direction-dependent differential structures and tensors and summarise an explicit construction of the space of directions at spatial infinity. We analyse the solutions of the electric Einstein equations at spatial infinity in \cref{sec:wave-hyp}. In \cref{sec:CD} we compare our formula for the Spi-supermomenta with the expression derived by Comp\`ere and Dehouck in \cite{CD}. In \cref{sec:BS}, we relate our covariant geometric construction to the approaches based on Bondi-Sachs and Beig-Schmidt coordinates.


\hr

We use an abstract index notation with indices \(a,b,c,\ldots\) for tensor fields. Quantities defined on the physical spacetime will be denoted by a ``hat'', while the ones on the conformally-completed unphysical spacetime are without the ``hat'' e.g. \(\hat g_{ab}\) is the physical metric while \(g_{ab}\) is the unphysical metric on the conformal-completion. We will raise and lower indices on tensors with \(g_{ab}\) and explicitly write out \(\hat g_{ab}\) when used to do so. We denote directions at \(i^0\) by an overhead arrow e.g. \(\vec N\) denotes directions which are either null or spatial while \(\vec\eta\) denotes spatial directions. Regular direction-dependent limits of tensor fields will be denoted by a boldface symbol e.g. \(\dd C_{abcd}(\vec\eta)\) is the limit of the (rescaled) unphysical Weyl tensor along spatial directions at \(i^0\). We collect our conventions on the orientations of the normals defined on various manifolds in \cref{tab:orientation}.

\begin{table}[!h]
\centering
\begin{tabular}{|l|l|}
	\hline
	 Normal vector field & Orientation \\ \hline
     \(n^a\) & null and future-pointing at \(\scri^+\), past-pointing at \(\scri^-\) \\
     \(l^a\) & null and future-pointing at \(\scri^+\), past-pointing at \(\scri^-\) \\
     \(\dd\eta^a\) & spatial and inward-pointing at \(i^0\) \\
     \(\dd u^a\) & timelike and future-pointing at some cross-section \(S\) of \(\hyp\) \\
     \(\dd\Sigma^{-1}\dd U^a\) & timelike and future-pointing at \(\nulls^+\), past-pointing at \(\nulls^-\) on \(\cyl\) \\
	\hline
\end{tabular}
\caption{Conventions for orientation of normals}
\label{tab:orientation}
\end{table}

\section{Asymptotic-flatness at null and spatial infinity: Ashtekar-Hansen structure}
\label{sec:AH}

We define spacetimes which are asymptotically-flat at null and spatial infinity using an Ashtekar-Hansen structure \cite{AH, Ash-in-Held} as follows. 

\begin{definition}[Ashtekar-Hansen structure \cite{Ash-in-Held}]\label{def:AH}
	A \emph{physical} spacetime \((\hat M, \hat g_{ab})\) has an \emph{Ashtekar-Hansen structure} if there exists another \emph{unphysical} spacetime \((M, g_{ab})\), such that
\begin{condlist}
	\item \(M\) is \(C^\infty\) everywhere except at a point \(i^0\) where it is \(C^{>1}\),
	\item the metric \(g_{ab}\) is \(C^\infty\) on \(M-i^0\), and \(C^0\) at \(i^0\) and \(C^{>0}\) along spatial directions at \(i^0\)
	\item there is an embedding of \(\hat M\) into \(M\) such that \(\bar J(i^0) = M - \hat M\),
	\item there exists a function \(\Omega\) on \(M\), which is \(C^\infty\) on \(M-i^0\) and \(C^2\) at \(i^0\) so that \(g_{ab} = \Omega^2 \hat g_{ab}\) on \(\hat M\) and
		\begin{condlist}
			\item \(\Omega = 0\) on \(\dot J(i^0)\) 
			\item \(\nabla_a \Omega \neq 0\) on \(\scri\)
			\item at \(i^0\), \(\nabla_a \Omega = 0\), \(\nabla_a \nabla_b \Omega = 2 g_{ab}\) \label{cond:Omega-at-i0}
		\end{condlist}
	\item There exists a neighbourhood \(N\) of \(\dot J(i^0)\) such that \((N, g_{ab})\) is  strongly causal and time orientable, and in \(N \inter \hat M\) the physical metric \(\hat g_{ab}\) satisfies the vacuum Einstein equation \(\hat R_{ab} = 0\),
	\item The space of integral curves of \(n^a = g^{ab}\nabla_b \Omega\) on \(\dot J(i^0)\) is diffeomorphic to the space of null directions at \(i^0\). \label{cond:int-curves}
    \item The vector field \(\varpi^{-1} n^a\) is complete on \(\scri\) for any smooth function \(\varpi\) on \(M - i^0\) such that \(\varpi > 0\) on \(\hat M \union \scri\) and \(\nabla_a(\varpi^4 n^a) = 0\) on \(\scri\). \label{cond:complete}
    \end{condlist}
\end{definition}
In the above we have used the following the notation for causal structures from \cite{Hawking-Ellis}: \(J(i^0)\) is the causal future of a point \(i^0\) in \(M\), \(\bar J(i^0)\) is its closure, \(\dot J(i^0) \) is its boundary and \(\scri \defn \dot J(i^0) - i^0\). We also use the definition and notation for direction-dependent tensors from \cref{sec:dd}. One can also use much weaker differentiability requirements on the unphysical metric \(g_{ab}\) (as discussed on \cref{sec:disc}), but we choose not to do so for simplicity.

The physical role of the conditions in \cref{def:AH} are explained in \cite{Ash-in-Held}. In particular, these conditions  imply that
\begin{enumerate*}[label=(\roman*)]
	\item The point \(i^0\) is spacelike related to all points in the physical spacetime \(\hat M\), and represents \emph{spatial infinity}.
	\item \(\scri \defn \dot J(i^0) - i^0\) consists of two disconnected pieces --- the future piece \(\scri^+\) and the past piece \(\scri^-\) --- which are both smooth null submanifolds of \(M\), representing future and past \emph{null infinities}, respectively.
\end{enumerate*}
Note that the metric \(g_{ab}\) is only \(C^{>0}\) at \(i^0\) along spatial directions, that is, the metric is continuous but the metric connection (or Christoffel symbols in some \(C^{>1}\) coordinate chart, see \cref{sec:dd}) is allowed to have limits which depend on the direction of approach to \(i^0\). As mentioned in the Introduction this low differentiability structure is necessary to accomodate spacetimes with non-vanishing ADM mass. Note that the unphysical metric is only required to be \(C^0\) approaching \(i^0\) along null directions; later we will impose additional regularity conditions (\cref{def:null-regular}) which ensure that the flux of all supermomenta through \(\scri\) is finite.\\

For a given physical spacetime \((\hat M, \hat g_{ab})\), the choice of an Ashtekar-Hansen structure is not unique. There is an ambiguity in the choice of the \(C^{>1}\) differential structure at \(i^0\) given by a \(4\)-parameter family of \emph{logarithmic translations} which simultaneously change the \(C^{>1}\)-structure and the conformal factor at \(i^0\) \cite{Berg, Ash-log, Chr-log} (see \cref{rem:log-trans}). Given a choice of the \(C^{>1}\)-structure the additional ambiguity in the choice of the conformal factor \(\Omega\) is as follows.
\begin{remark}[Freedom in the conformal factor \cite{AH,Ash-in-Held}]
\label{rem:freedom-Omega}
The freedom in the choice of the conformal factor in \cref{def:AH} is given by \(\Omega \mapsto \omega\Omega\) where the function \(\omega\) satisfies
\begin{propertylist}
    \item \(\omega > 0\) on \(M\) 
    \item \(\omega\) is smooth on \(M - i^0\)
    \item \(\omega\) is \(C^{>0}\) in spatial directions at \(i^0\) and \(\omega \vert_{i^0} = 1\)
\end{propertylist}
We say that a tensor field \(T^{a\ldots}{}_{b\ldots}\) has a \emph{conformal weight} \(w\) if under the above change of conformal factor it transforms as
\be\label{eq:conf-wt-defn}
    T^{a\ldots}{}_{b\ldots} \mapsto \omega^w~ T^{a\ldots}{}_{b\ldots}
\ee
For instance the unphysical metric \(g_{ab}\) has conformal weight \(w = 2\). 
\end{remark}

In the following we will work with a fixed the unphysical spacetime given by some choice of the \(C^{>1}\)-structure at \(i^0\) and some choice of conformal factor \(\Omega\). In the end, we argue that our results are independent of these choices (see \cref{rem:change-Sigma,rem:lor-inv,rem:log-trans-inv}). Note that, all spacetimes satisfying \cref{def:AH} have the same metric at \(i^0\), that is, the unphysical metric \(g_{ab}\) at \(i^0\) is \emph{universal} (isometric to the Minkowski metric at \(i^0\)) and cannot even be further conformally-rescaled (since \(\omega\vert_{i^0} = 1\)) \cite{AH}.

Using the conformal transformation relating the unphysical Ricci tensor \(R_{ab}\) to the physical Ricci tensor \(\hat R_{ab}\) (see Appendix~D \cite{Wald-book}), the vacuum Einstein equation \(\hat R_{ab} = 0\) can be written as
\be\label{eq:EE}
    S_{ab} = - 2 \Omega^{-1} \nabla_{(a} n_{b)} + \Omega^{-2} n^c n_c g_{ab}
\ee
where \(S_{ab}\) is given by
\be\label{eq:S-defn}
    S_{ab} \defn R_{ab} - \tfrac{1}{6} R g_{ab}
\ee
Further, the Bianchi identity \(\nabla_{[a} R_{bc]de} = 0\) on the uphysical Riemann tensor along with \cref{eq:EE} gives the following equations for the unphysical Weyl tensor \(C_{abcd}\) (see \cite{Geroch-asymp} for details)
\begin{subequations}\label{eq:Bianchi-unphys}\begin{align}
    \nabla_{[a} (\Omega^{-1} C_{bc]de}) = 0 \label{eq:curl-weyl}\\
    \nabla^d C_{abcd} = - \nabla_{[a} S_{b]c} \label{eq:Weyl-S}
\end{align}\end{subequations}

\hr

On \(\scri\), let us introduce the function
\be\label{eq:Phi-defn}
    \Phi \defn \tfrac{1}{4} \nabla_a n^a\vert_\scri \eqsp \Phi\vert_{i^0} = 2 
\ee
where the second condition follows from \cref{cond:Omega-at-i0}. Since \(S_{ab}\) is smooth at \(\scri\), by the assumptions in \cref{def:AH}, \cref{eq:EE} implies
\be\label{eq:n-Phi}
    \lim_{\to \scri} \Omega^{-1}n^a n_a = 2 \Phi \eqsp \nabla_a n_b \vert_\scri = \Phi g_{ab} 
\ee 
that is, the vector field \(n^a\) is a null geodesic generator of \(\scri^\pm \cong \bb R \times \bb S^2\), which is future/past pointing on \(\scri^\pm\) respectively.

 Denote by \(q_{ab}\) the pullback of \(g_{ab}\) to \(\scri\). This defines a degenerate metric on \(\scri\) with \(q_{ab} n^b = 0\). It is convenient to introduce a foliation of \(\scri\) by a family of cross-sections diffeomorphic to \(\bb S^2\). The pullback of \(q_{ab}\) to any cross-section \(S\) defines a Riemannian metric on \(S\). Then, for any choice of foliation, there is a unique \emph{auxilliary normal} vector field \(l^a\) at \(\scri\) such that
\be\label{eq:l-props}
    l^a l_a = 0 \eqsp l^a n_a = -1 \eqsp q_{ab}l^b = 0
\ee
In our conventions, \(l^a\) is future/past pointing at \(\scri^\pm\), respectively (\cref{tab:orientation}). We further have
\be\label{eq:null-fields}
    q_{ab} = g_{ab} + 2 n_{(a} l_{b)} \eqsp \varepsilon_{abc} = l^d \varepsilon_{dabc} \eqsp \varepsilon_{ab} = n^c \varepsilon_{cab}
\ee
where \(\varepsilon_{abc}\) defines a volume element on \(\scri\) and \(\varepsilon_{ab}\) is the area element on any cross-section \(S\) of the foliation. Evaluating the pullback of \(\Lie_n g_{ab}\) and using \cref{eq:n-Phi} we have on \(\scri\)
\be\label{eq:Lie-n-q}
    \Lie_n q_{ab} = 2 \Phi q_{ab}
\ee
that is, \(\Phi\) measures the expansion of the chosen cross-sections of \(\scri\) along the null generator \(n^a\) while, their shear and twist identically vanishes.

Let
\be\label{eq:tau-defn}
    \tau_a \defn q_a{}^c n^b \nabla_b l_c
\ee
so we have \(n^b \nabla_b l_a = \tau_a - \Phi l_a\) i.e., \(\tau_a\) represents change in the direction of \(l_a\) along the null generators of \(n^a\). The shear of the auxilliary normal \(l^a\) on the cross-sections \(S\) of the foliation is defined by
\be\label{eq:sigma-defn}
    \sigma_{ab} \defn (q_a{}^c q_b{}^d - \tfrac{1}{2} q_{ab} q^{cd} ) \nabla_c l_d 
\ee
while the twist \(\varepsilon^{ab}\nabla_a l_b\) vanishes on account of \(l_a\) being normal to the cross-sections. We will not require the expansion of \(l^a\) in our analysis.

For any smooth \(v_a\) satisfying \(n^a v_a = l^a v_a = 0\) on \(\scri\) we define the derivative \(\ms D_a\) on the cross-sections by
\be
    \ms D_a v_b \defn q_a{}^c q_b{}^d \nabla_c v_d
\ee
It is easily verified that \(\ms D_a q_{bc} = 0\), i.e., \(\ms D_a\) is the metric-compatible covariant derivative on cross-sections of \(\scri\). We also note the following identity for integration-by-parts on any cross-section \(S\) of \(\scri\)
\be\label{eq:IBP}\begin{split}
    \int_S \varepsilon_2 ~ f \ms D^a v_a = - \int_S \varepsilon_2 ~ (\nabla_a f q^{ab} + f \nabla_a q^{ab}) v_b 
    = - \int_S \varepsilon _2 ~ (\ms D_a + \tau_a) f v^a
\end{split}\ee
where we have discarded a boundary term on \(S \cong \bb S^2\), used \cref{eq:null-fields,eq:tau-defn} and that \(v_a\) is orthogonal to both \(n^a\) and \(l^a\). The generalisation to arbitrary tensors on \(\scri\) that are orthogonal to both \(n^a\) and \(l^a\) is immediate.

On \(\scri\), with the choice of foliation of \(\scri\) held fixed, we have the following conformal weights (\cref{eq:conf-wt-defn})\footnote{Note that the normal \(n_a\) also transforms away from \(\scri\) as \(n_a \mapsto \omega n_a + \Omega \nabla_a \omega\).}
\be\label{eq:conf-trans1}
    (n_a, l_a) : w = 1 \eqsp (n^a, l^a) : w = -1 \eqsp q_{ab} : w =2 \eqsp \sigma_{ab} : w = 1
\ee
while \(\Phi\) and \(\tau_a\) are not conformally-weighted but transform as
\be\label{eq:conf-trans}
    \Phi \mapsto \omega^{-1} (\Phi + \Lie_n \ln \omega) \eqsp \tau_a \mapsto \tau_a + \ms D_a \ln\omega 
\ee

\begin{remark}[Choices of conformal factor]
\label{rem:conf-choices}
It has been conventional to choose the conformal factor \(\Omega\) so that the \emph{Bondi condition} holds 
\be
    \nabla_a \nabla_b \Omega \vert_\scri = 0
\ee
This can be seen to be equivalent to the condition that \(\Phi = 0\) at \(\scri\). However, such a choice of conformal factor violates \cref{cond:Omega-at-i0} i.e., \(\Phi\vert_{i^0} = 2\), and is ill-behaved at \(i^0\). It can be verified that the function \(\varpi\) in \cref{cond:complete} used to define a complete divergence-free normal \(\varpi^{-1}n^a\) cannot be used as a conformal-rescaling at \(i^0\) since \(\varpi\) will diverge at \(i^0\) (see footnote 2, \S~11.1 \cite{Wald-book} and \cref{sec:BS}). In particular, the unphysical metric in the Bondi conformal frame and the corresponding Bondi-Sachs coordinates are ill-behaved near \(i^0\). It is however always possible to choose the conformal factor so that \(\Phi = 2\) on \(\scri\) in some neighbourhood around \(i^0\). This choice was made, for instance, in \cite{Ash-Mag-Ash, HL-GR-matching} and simplifies many of the subsequent computations. We prefer to keep the choice of conformal factor arbitrary, subject to \cref{rem:freedom-Omega}, so that the conformal invariance of our result can be easily verified.
\end{remark}

\hr 

At spatial infinity represented by a single point \(i^0\) in \(M\), the gravitational fields of interest, in general, only admit \emph{direction-dependent} limits, and hence it is rather awkward to study such fields directly at \(i^0\). Instead of working at the point \(i^0\) one works on the space of directions at \(i^0\) i.e., a \emph{blowup} of \(i^0\) (see \cite{Harris}). The fields which have regular direction-dependent limits to \(i^0\) (as defined in \cref{sec:dd}) induce smooth tensor fields on the space of directions. The space of spatial directions at \(i^0\) was constructed in \cite{AH}, we review the aspects of this construction needed in our analysis below. A different, but related, blowup of \(i^0\) which includes the null directions at \(i^0\) was constructed in \cite{KP-EM-match} (summarised in \cref{sec:cyl}) and is more useful for relating the gravitational fields and symmetries at spatial infinity to those on null infinity.

\begin{figure}[h!]
	\centering
	\includegraphics[width=0.3\textwidth]{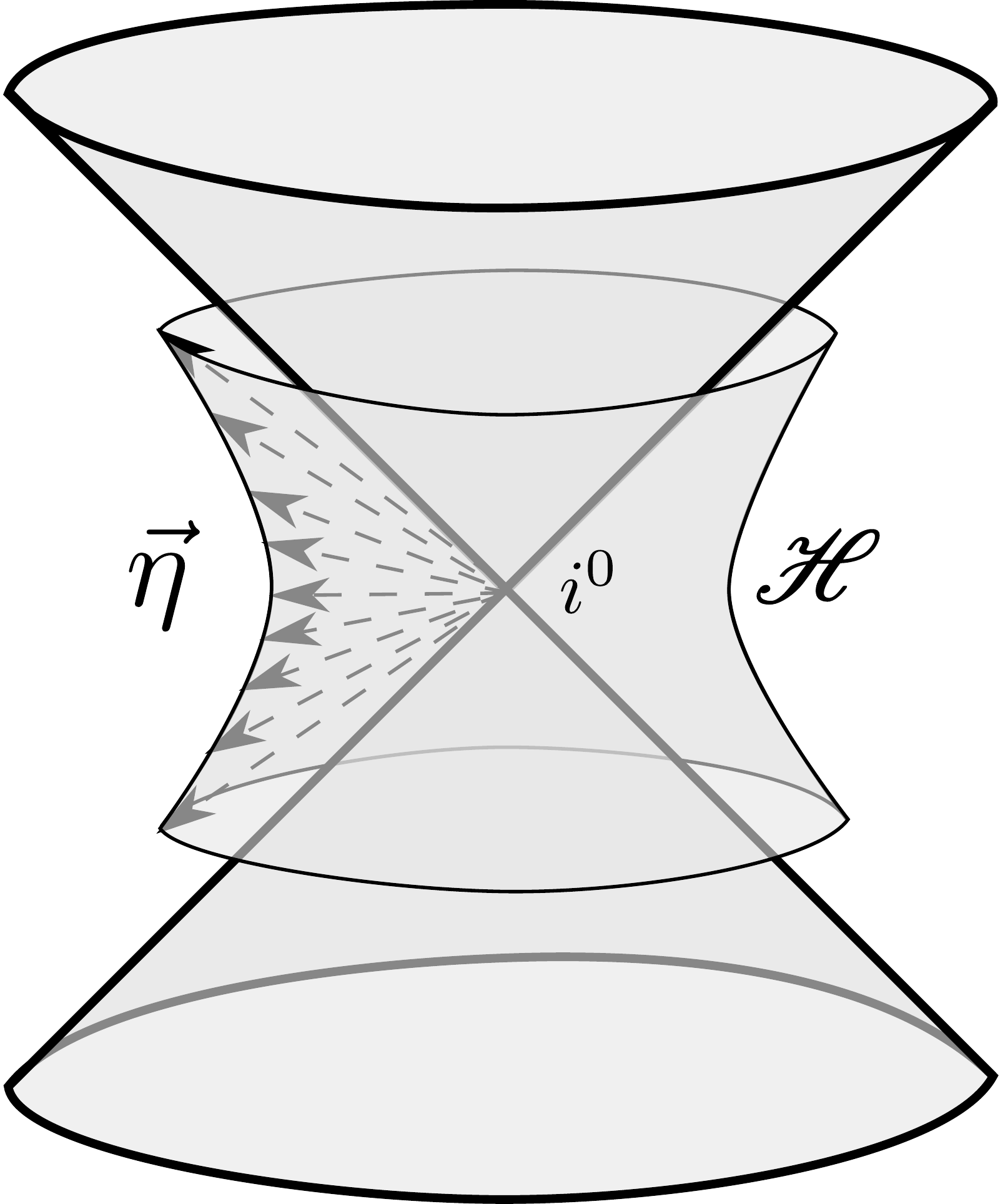}
	\caption{The (non-compact) unit-hyperboloid \(\hyp\) in \(Ti^0\) representing spatial directions \(\vec\eta\) at \(i^0\).}
    \label{fig:hyperboloid}
\end{figure}

Along spatial directions \(\nabla_a \Omh\) is \(C^{>-1}\) at \(i^0\) and 
\be\label{eq:eta-defn}
    \dd\eta^a \defn \lim_{\to i^0} \nabla^a \Omh
\ee
determines a \(C^{>-1}\) spatial unit vector field at \(i^0\) representing the spatial directions \(\vec\eta\) at \(i^0\). The space of directions \(\vec\eta\) in \(Ti^0\) is a unit-hyperboloid \(\hyp\) depicted in \cref{fig:hyperboloid}. 

If \(T^{a \ldots}{}_{b \ldots}\) is a \(C^{>-1}\) tensor field at \(i^0\) in spatial directions then, \(\lim_{\to i^0} T^{a \ldots}{}_{b \ldots} = \dd T^{a \ldots}{}_{b \ldots}(\vec\eta)\) is a smooth tensor field on \(\hyp\). Further, the derivatives of \(\dd T^{a \ldots}{}_{b \ldots}(\vec\eta)\) to all orders with respect to the direction \(\vec\eta\) satisfy\footnote{The factors of \(\Omh\) on the right-hand-side of \cref{eq:dd-der-spatial} convert between \(\nabla_a\) and the derivatives with respect to the directions \cite{Ash-in-Held,Geroch-asymp}.}
\be\label{eq:dd-der-spatial}
    \dd \partial_c \cdots \dd \partial_d \dd T^{a \ldots}{}_{b \ldots}(\vec\eta) = \lim_{\to i^0} \Omh \nabla_c \cdots \Omh \nabla_d T^{a \ldots}{}_{b \ldots}
\ee
where \(\dd \partial_a\) is the derivative with respect to the directions \(\vec \eta\) defined by 
\be\label{eq:dd-derivative-spatial}\begin{split}
    \dd v^c \dd \partial_c \dd T^{a \ldots}{}_{b \ldots}(\vec\eta) & \defn \lim_{\epsilon \to 0} \frac{1}{\epsilon} \big[ \dd T^{a \ldots}{}_{b \ldots}(\vec\eta + \epsilon \vec v) - \dd T^{a \ldots}{}_{b \ldots}(\vec\eta) \big] \quad \text{for all } \dd v^a \in T\hyp \\
    \dd \eta^c \dd \partial_c \dd T^{a \ldots}{}_{b \ldots}(\vec\eta) & \defn 0
\end{split}\ee
It can be checked that, along spatial directions, \cref{eq:dd-der-spatial} is equivalent to the definition \cref{eq:dd-derivative} given in terms of a \(C^{>1}\) coordinate chart.

The metric \(\dd h_{ab}\) induced on \(\hyp\) by the universal metric \(\dd g_{ab}\) at \(i^0\), satisfies
\be\label{eq:d-eta-h}
    \dd h_{ab} \defn \dd g_{ab} - \dd \eta_a \dd \eta_b = \dd \partial_a \dd \eta_b
\ee
Further, if \(\dd T^{a \ldots}{}_{b \ldots}(\vec\eta)\) is orthogonal to \(\dd\eta^a\) in all its indices then it defines a tensor field \(\dd T^{a \ldots}{}_{b \ldots}\) intrinsic to \(\hyp\). In this case, projecting \emph{all} the indices in \cref{eq:dd-der-spatial} using \(\dd h_{ab}\) to defines a derivative operator \(\dd D_a\) intrinsic to \(\hyp\) which is also the covariant derivative operator associated to \(\dd h_{ab}\).\footnote{This follows from \cref{eq:d-eta-h}, and \(\dd\partial_c \dd g_{ab} = 0\) since \(\dd g_{ab}\) is direction-independent at \(i^0\).} We also define
\be\label{eq:volume-hyp}
    \dd\varepsilon_{abc} \defn - \dd\eta^d \dd\varepsilon_{dabc} \eqsp \dd\varepsilon_{ab} \defn \dd u^c \dd\varepsilon_{cab}
\ee
where \(\dd\varepsilon_{abcd}\) is volume element at \(i^0\) corresponding to the metric \(\dd g_{ab}\), \(\dd\varepsilon_{abc}\) is the induced volume element on \(\hyp\), and \(\dd\varepsilon_{ab}\) is the induced area element on some cross-section \(S\) of \(\hyp\) with a future-pointing timelike normal \(\dd u^a\) such that \(\dd h_{ab} \dd u^a \dd u^b = -1\). 

We note that \(\hyp\) admits a reflection isometry as follows. On the unit-hyperboloid we can introduce coordinates \((\tau, \theta^A)\) --- where \(\tau \in (-\infty,\infty)\), and \(\theta^A = (\theta,\phi )\) are the standard coordinates on \(\bb S^2\) with \(\theta \in [0,\pi]\) and  \(\phi \in [0 ,2\pi)\) --- so that the metric on \(\hyp\) is
\be\label{eq:h-hyp-tau}
	\dd h_{ab} \equiv - d\tau^2 + \cosh^2\tau (d\theta^2 + \sin^2\theta d\phi^2 )
\ee
The metric \(\dd h_{ab}\) has a reflection isometry \(\Upsilon\) 
\be\label{eq:reflection-hyp}\begin{split}
    \Upsilon &: \hyp \to \hyp : (\tau, \theta^A) \mapsto (-\tau, -\theta^A) \\
    \text{with } \Upsilon &\circ \dd h_{ab} = \dd h_{ab}
\end{split}\ee
where \(\theta^A = (\theta, \phi) \mapsto - \theta^A = (\pi - \theta, \phi\pm \pi)\) is the antipodal reflection on \(\bb S^2\) --- the sign is chosen so that \(\phi \pm \pi \in [0,2\pi)\). We have used \(\Upsilon \circ\) to denote the natural action of the reflection map \(\Upsilon\) on tensor fields on \(\hyp\).

\subsection{The space \(\cyl\) of null and spatial directions at \(i^0\)}
\label{sec:cyl}

\begin{figure}[h!]
	\centering
	\includegraphics[width=0.3\textwidth]{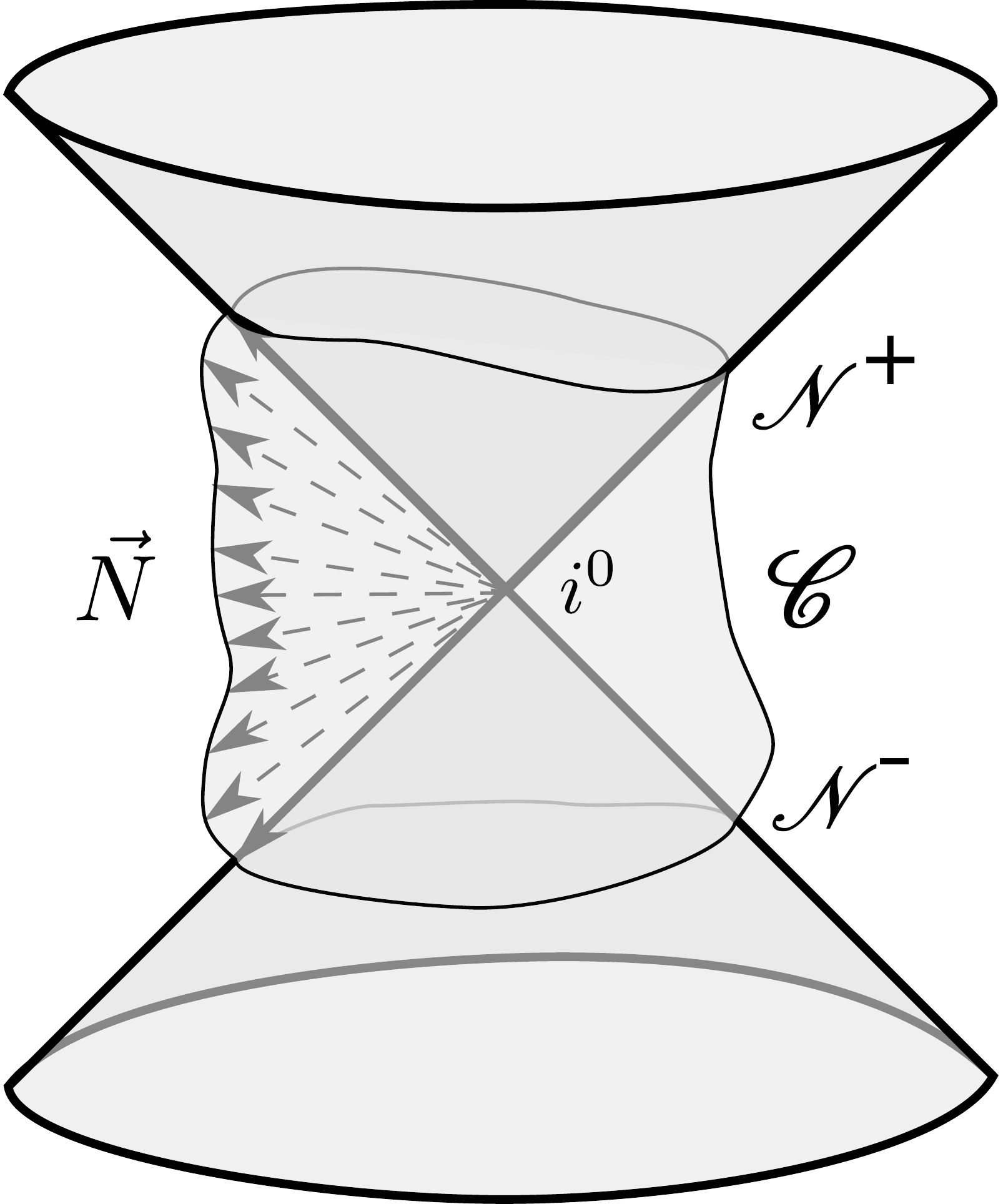}
	\caption{The space \(\cyl\) of null and spatial directions \(\vec N\) at \(i^0\). The boundaries \(\nulls^\pm \cong \bb S^2\), diffeomorphic to the space of generators of \(\scri^\pm\) respectively, represent the space of null directions. \(\cyl \setminus \nulls^\pm\) is the space of \emph{rescaled} spatial directions conformally diffeomorphic to the unit-hyperboloid \(\hyp\). \(\cyl\) depends on the choice of the rescaling function \(\Sigma\) (defined below) and \emph{need not} be a cylinder of unit radius in \(Ti^0\) --- we have drawn a ``wiggly'' cylinder to emphasise this.} 
    \label{fig:cylinder}
\end{figure}

For the matching problem we are interested in analysing the gravitational fields and supertranslations in the null directions along \(\scri^\pm\) at \(i^0\). The hyperboloid \(\hyp\) is not well-suited for this as the null directions correspond to ``points at infinity'' on \(\hyp\). Thus, we use a different blowup \(\cyl\) of \(i^0\), constructed in \cite{KP-EM-match}, which includes both null and spatial directions. The basic strategy of the construction is as follows.
\begin{enumerate*}
    \item We rescale \(n^a\) so that the rescaled vector field is non-vanishing and represents ``good'' null directions at \(i^0\).
    \item We also conformally-complete \(\hyp\) to get a new manifold whose boundaries represent the points in the infinite future or past along \(\hyp\).
    \item Then, we can identify the null directions given by the rescaled \(n^a\) with the boundaries of the conformal-completion of \(\hyp\) in a ``sufficiently smooth'' way, to get the new manifold \(\cyl\).
\end{enumerate*}
The final picture obtained is depicted in \cref{fig:cylinder}.

To implement the construction of \(\cyl\) described above, we work in a neighbourhood of \(i^0\) in \(M\), and use \(M\) to mean such a neighbourhood unless otherwise specified. In \(M\), we define a \emph{rescaling function} \(\Sigma\) as follows:

\begin{definition}[Rescaling function \(\Sigma\)]
\label{def:Sigma}
Let \(\Sigma\) be a function in \(M\) such that 
\begin{condlist}
    \item \(\Sigma^{-1} > 0\) is smooth on \(M - i^0\)
    \item \(\Sigma^{-1}\) is \(C^{>0}\) at \(i^0\) in both null and spatial directions,
	\item \(\Sigma^{-1}\vert_{i^0} = 0\), \(\lim_{\to i^0} \nabla_a \Sigma^{-1} \neq 0\) and \label{cond:Sigma-i0}
    \item \(\Sigma \Lie_n \Sigma^{-1} = 2\) at \(i^0\) and on \(\scri\) \label{cond:Sigma-choice}
\end{condlist}
\end{definition}

\begin{remark}[Freedom in the rescaling function]
\label{rem:freedom-Sigma}
If \(\Sigma\) is a choice of rescaling function for the conformal factor \(\Omega\) then \(\sigma\Sigma\) is a choice for the conformal factor \(\omega\Omega\) (where \(\omega\) satisfies the conditions in \cref{rem:freedom-Omega}) if
\begin{propertylist}
    \item \(\sigma > 0\) in \(M\)
    \item \(\sigma\) is smooth on \(M - i^0\)
    \item \(\sigma\) is \(C^{>-1}\) at \(i^0\) in both null and spatial directions
    \item \(\Lie_n \ln\sigma\vert_\scri = 2(1-\omega)\vert_\scri\)
\end{propertylist} 
Similar, to the conformal weight (\cref{eq:conf-wt-defn}), we say that a tensor field \(T^{a\ldots}{}_{b\ldots}\) has a \emph{rescaling weight} \(r\) if under the above change of rescaling function (with fixed conformal factor, i.e. \(\omega = 1\)) it transforms as
\be\label{eq:rescaling-wt-defn}
    T^{a\ldots}{}_{b\ldots} \mapsto \sigma^r~ T^{a\ldots}{}_{b\ldots}
\ee
\end{remark}

The existence of such a rescaling function was shown in Appendix~B \cite{KP-EM-match} (summarised in \cref{sec:dd} below). We emphasise that the rescaling by \(\Sigma\) is \emph{not} an alternative choice of the conformal factor \(\Omega\), in particular the unphysical metric \(g_{ab}\) is not rescaled by \(\Sigma\). Any choice of the rescaling function \(\Sigma\) allows us to construct certain regular fields near \(i^0\) which will be useful in our analysis. We list below their essential properties which can be verified following in Appendix~B \cite{KP-EM-match}.

Since \(\Sigma^{-1}\vert_{i^0} = 0\) and \(\Sigma^{-1}\) is \(C^{>0}\), there exists a function \(\dd\Sigma(\vec\eta)\), which is \(C^{>-1}\) along spatial directions, such that
\be\label{eq:dd-Sigma}
    \dd\Sigma^{-1}(\vec\eta) = \lim_{\to i^0} (\Omh \Sigma)^{-1}
\ee

\paragraph*{Rescaled normal \(N^a\) and the space \(\cyl\) of null and spatial directions:}
The rescaled vector field (the factor of half in definition of \(N^a\) is for later convenience)
\be\label{eq:N-defn} 
N^a \defn \tfrac{1}{2} \Sigma n^a = \tfrac{1}{2} \Sigma \nabla^a \Omega
\ee
is \(C^{>-1}\) at \(i^0\) such that \(\dd N^a = \lim_{\to i^0} N^a \neq 0\) in both null and spatial directions. Thus, along \(\scri\), we have \(\dd N^a\) as a direction-dependent null vector representing the null directions at \(i^0\) which are future/past directed along \(\scri^\pm\) respectively. Along spatial directions at \(i^0\) we have \(\dd N^a(\vec\eta) = \dd\Sigma(\vec\eta) \dd\eta^a \neq 0\) which represents the rescaled spatial directions at \(i^0\). The space of these directions \(\vec N\) can be represented by a cylinder \(\cyl\) with two boundaries \(\nulls^\pm \cong \bb S^2\) (as in \cref{fig:cylinder}). The boundaries \(\nulls^\pm\) represent the null directions along \(\scri^\pm\) respectively, while \(\cyl\setminus \nulls^\pm\) is the space of the rescaled spatial directions at \(i^0\).

\paragraph*{Rescaled auxilliary normal \(L^a\) and foliation of \(\scri\):}
Define a vector field \(L^a\) in \(M\) by
\be\label{eq:L-defn}
    L^a \defn - \nabla^a \Sigma^{-1} + \tfrac{1}{2}\nabla_b \Sigma^{-1} \nabla^b \Sigma^{-1} ~ N^a
\ee 
which is \(C^{>-1}\) at \(i^0\) and \(\lim_{\to i^0} L^a \neq 0\) in both null and spatial directions. Further, using \cref{eq:N-defn,cond:Sigma-choice}, we have
\be\label{eq:NL-LL}
    N^a L_a\vert_\scri = -1 \eqsp L^a L_a\vert_\scri = 0
\ee
The pullback to \(\scri\) of \(L_a\) equals the pullback of \(- \nabla_a \Sigma^{-1}\), thus \(L^a\) defines a rescaled auxilliary normal to the foliation of \(\scri\) by a family of cross-sections \(S_\Sigma\) with \(\Sigma^{-1} = \text{constant}\). From \cref{def:Sigma} and \cref{cond:int-curves}, the limiting cross-section \(S_\Sigma\) as \(\Sigma^{-1} \to 0\), is diffeomorphic to the space of null directions \(\nulls^\pm\). The auxilliary normal \(l^a\) to this foliation, satisfying \cref{eq:l-props}, is obtained by
\be\label{eq:l-defn}
    l^a \defn \tfrac{1}{2} \Sigma L^a
\ee
We also extend \(l^a\) into \(M\) by the above formula and \cref{eq:L-defn}. In such a foliation, we have (using \cref{eq:N-defn,eq:NL-LL})
\be\label{eq:N-n-Sigma}
    N^a\vert_\scri \equiv \partial_{\Sigma^{-1}} \eqsp n^a\vert_\scri \equiv 2 \Sigma^{-1} \partial_{\Sigma^{-1}}    
\ee
Further, we can show that the tensor \(\tau_a\) defined in \cref{eq:tau-defn} vanishes in this choice of foliation. From \cref{eq:L-defn,eq:l-defn} we can compute
\be\label{eq:tau-0}
    \tau_a = - \tfrac{1}{2} q_a{}^c n^b \nabla_b \nabla_c \ln \Sigma^{-1} = - \tfrac{1}{2} q_a{}^c n^b \nabla_c \nabla_b \ln \Sigma^{-1} = \tfrac{1}{2} \Phi \ms D_a \ln \Sigma^{-1} = 0
\ee
where we have used \cref{eq:n-Phi,cond:Sigma-choice} on \(\scri\). Finally, the vanishing of \(\tau_a\) follows since each cross-section of the foliation has \(\Sigma^{-1} = \text{constant}\).\footnote{Note that under a conformal transformation \(\tau_a\) changes according to \cref{eq:conf-trans}. But the rescaling function \(\Sigma\), and hence the chosen foliation by \(S_\Sigma\), also changes according to \cref{rem:freedom-Sigma}, so that the condition \(\tau_a = 0\) also holds in the new (and thus any) choice of conformal factor and choice of foliation of \(\scri\) by \(S_\Sigma\).}

\paragraph*{Conformal-completion of \(\hyp\):}
Let \(\dd\Sigma\) be the function induced on \(\hyp\) by \(\dd\Sigma(\vec\eta)\) (\cref{eq:dd-Sigma}). Let \((\tilde\hyp, \tilde{\dd h}_{ab}) \) be a conformal-completion of \((\hyp, \dd h_{ab})\) with metric  \(\tilde{\dd h}_{ab} = \dd\Sigma^2 \dd h_{ab}\). There exists a diffeomorphism from \(\tilde\hyp\) onto \(\cyl\) such that \(\hyp\) is mapped onto \(\cyl \setminus \nulls^\pm\) and \(\dd\Sigma\), as a function on \(\cyl \setminus \nulls^\pm\), extends smoothly to the boundaries \(\nulls^\pm\) where
\be
    \dd\Sigma\vert_{\nulls^\pm} = 0
\ee
Using \cref{eq:L-defn,eq:eta-defn,eq:d-eta-h,eq:dd-der-spatial}, the limit of \(L^a\) to \(i^0\) along spatial directions gives the direction-dependent vector field
\be
    \dd L^a(\vec\eta) = - \dd h^{ab} \dd D_b \dd\Sigma^{-1} + (\dd h^{bc} \dd D_b \dd\Sigma^{-1} \dd D_c \dd\Sigma^{-1}) \dd\eta^a
\ee
The projection of \(\dd L^a(\vec\eta)\) onto \(\hyp\) is the vector field
\be\label{eq:U-defn}
    \dd U^a \defn \dd h^a{}_b \dd L^b(\vec\eta) = \tilde{\dd h}^{ab} \dd D_b \dd \Sigma
\ee
Viewed as a vector field on \(\tilde\hyp\), and hence \(\cyl\), we have
\be\label{eq:U-on-N}
    \lim_{\to \nulls^\pm}\tilde{\dd h}_{ab} \dd U^a \dd U^b = -1 \eqsp \lim_{\to \nulls^\pm}\dd\Sigma^{-1} \dd U^a \neq 0 
\ee
Note that \(\dd\Sigma^{-1}\dd U^a\) is future/past directed at \(\nulls^\pm\) respectively (see \cref{tab:orientation}). Note, from \cref{eq:U-on-N}, the metric \(\tilde{\dd h}_{ab}\) is \emph{not} smooth at \(\nulls^\pm\) on \(\cyl\), but still provides a useful relation between \(\cyl\) and the conformal-completion of \(\hyp\).

\paragraph*{Metric on \(\nulls^\pm\):}
On \(\scri\) consider the rescaled metric
\be\label{eq:rescaled-q}
    \tilde q_{ab} \defn \Sigma^2 q_{ab}
\ee
Along the foliation \(S_\Sigma\) as \(\Sigma^{-1} \to 0\), \(\lim_{\to i^0} \tilde q_{ab}(\vec N)\) exists along null directions \(\vec N\) and defines a direction-dependent Riemannian metric \(\tilde{\dd q}_{ab}\) on the space of null directions \(\nulls^\pm\). Further, this metric coincides with the metric induced on \(\nulls^\pm\) by \(\tilde{\dd h}_{ab}\) on \(\cyl\), that is,
\be
    \tilde{\dd q}_{ab} = \lim_{\to \nulls^\pm} (\tilde{\dd h}_{ab} + \dd D_a \dd\Sigma \dd D_b \dd\Sigma)
\ee
Similarly, the rescaled area element \(\tilde\varepsilon_{ab} \defn \Sigma^2 \varepsilon_{ab}\) on the foliation \(S_\Sigma\) induces an area element \(\tilde{\dd\varepsilon}_{ab}\) on \(\nulls^\pm\) such that
\be\label{eq:nulls-area}
    \tilde{\dd\varepsilon}_{ab} = \lim_{\to \nulls^\pm} \dd U^c \tilde{\dd\varepsilon}_{cab} 
\ee
where \(\tilde{\dd\varepsilon}_{abc} \defn \dd\Sigma^3 \dd\varepsilon_{abc}\) is the volume element on \(\cyl\) defined by the metric \(\tilde{\dd h}_{ab} = \dd\Sigma^2 \dd h_{ab}\).

\paragraph*{Reflection conformal isometry of \(\cyl\):}
The reflection isometry \(\Upsilon\) of \(\hyp\) (see \cref{eq:reflection-hyp}) extends to a reflection conformal isometry of \(\cyl\) i.e., there exists a reflection map \(\Upsilon : \cyl \to \cyl\) and a smooth function \(\dd\varsigma > 0\) on \(\cyl\) such that
\be\label{eq:reflection-cyl}
    \Upsilon \circ \tilde{\dd h}_{ab} = \dd\varsigma^2 \tilde{\dd h}_{ab}
\ee
where \(\dd\varsigma = \dd \Sigma^{-1} (\Upsilon \circ \dd \Sigma) \) accounts for the fact that choice of rescaling function need not be invariant under the reflection \(\Upsilon\). Further, under this reflection map we also have \(\Upsilon : \nulls^- \to \nulls^+\), such that
\be\label{eq:reflection-cyl-boundary}\begin{split}
     \Upsilon \circ (\dd\Sigma^{-1} \dd U^a)\vert_{\nulls^-} & = - \dd\varsigma^{-2} \dd\Sigma^{-1} \dd U^a \vert_{\nulls^+} \\
    \Upsilon \circ \tilde{\dd \varepsilon}_{ab}\vert_{\nulls^-} & = - \dd\varsigma^2 \tilde{\dd \varepsilon}_{ab}\vert_{\nulls^+}
\end{split}\ee
where the negative signs on the right-hand-side are due to our orientation conventions (\cref{tab:orientation}). From \cref{cond:int-curves}, \(\nulls^\pm\) are diffeomorphic to the space of null generators of \(\scri^\pm\), and thus the reflection map \(\Upsilon : \nulls^- \to \nulls^+\) provides an antipodal map between the null generators of \(\scri^-\) and those of \(\scri^+\). This mapping between the generators of \(\scri^\pm\) is singled out by the fact that it is a symmetry of those solutions to the Einstein equation on \(\cyl\) which smoothly extend to \(\nulls^\pm\) (see \cref{sec:wave-hyp}).

\paragraph*{The abstract manifold \(\cyl\) with a conformal-class of metrics:}
If we choose a different rescaling function \(\Sigma' = \sigma \Sigma\) with \(\sigma\) satisfying the conditions in \cref{rem:freedom-Sigma}, we get a different space \(\cyl'\) of directions \(\vec N' = \dd\sigma(\vec N) \vec N\) at \(i^0\), where \(\dd\sigma(\vec N) = \lim_{\to i^0} \sigma\) along the directions \(\vec N\). This new space \(\cyl'\) is naturally diffeomorphic to \(\cyl\) under the above mapping of the directions with the metric \(\tilde{\dd h}_{ab}' = \dd \sigma^2 \tilde{\dd h}_{ab}\). Thus, we consider \(\cyl\) as an abstract manifold with this conformal-class of metrics --- note that here the conformal-class corresponds to a change of rescaling function and \emph{not} the conformal factor \(\Omega\). This point of view will be useful to show that our results are Lorentz-invariant; \cref{rem:lor-inv}. The transformation of the various fields defined above can be computed directly from the defining equations.

\paragraph*{\(C^{>-1}\) functions on \(\cyl\):}
Consider any function \(f\) which is smooth at \(\scri\) and \(C^{>-1}\) at \(i^0\) in \emph{both} null and spatial directions. Then along null directions, \(\lim_{\to i^0}f\vert_{\scri^\pm}\) induces a smooth function \(\dd f^\pm\) on \(\nulls^\pm\). Similarly, along spatial directions, \(\lim_{\to i^0} f \) induces a smooth function \(\dd f\) on \(\hyp\). Using the diffeomorphism between \(\hyp\) and \(\cyl \setminus \nulls^\pm\), we can consider \(\dd f\) as a smooth function on \(\cyl\setminus \nulls^\pm\). Since, \(f\) is \(C^{>-1}\) along both null and spatial directions, the function \(\dd f\) extends to \(\nulls^\pm\) as a smooth function, and on \(\nulls^\pm\) satisfies 
\be
    \dd f^\pm = \dd f \vert_{\nulls^\pm}
\ee
That is, for functions which are \(C^{>-1}\) in both null and spatial directions, the fields induced on \(\nulls^\pm\) by, first taking the limit to \(\scri^\pm\) and then to \(i^0\), or by, first taking the limit to \(i^0\) in spatial directions and then to the space of null directions \(\nulls^\pm\), coincide. Thus, such functions are continuous at the space of null directions \(\nulls^\pm\) when going from \(\cyl\) to \(\scri^\pm\).

\section{Supertranslations and supermomentum in general relativity}
\label{sec:super-trans}

The well-known expressions for the BMS-supertranslations and the corresponding supermomenta hold only in choices of the conformal factor which satisfy the Bondi condition on \(\scri\) which, as mentioned above (\cref{rem:conf-choices}), are incompatible with the \cref{cond:Omega-at-i0}. Thus, we first need to extend these expressions to arbitrary choices of the conformal factor. The computation of these expressions is most easily done in Geroch-Held-Penrose (GHP) formalism, and so we defer the details to \cref{sec:non-Bondi}. The relation to the expressions existing in the literature is given in \cref{rem:prev-rel}.

By the peeling theorem, we have \(C_{abcd} = 0\) at \(\scri\), and thus \(\Omega^{-1}C_{abcd}\) admits a limit to \(\scri\) (see Theorem~11 \cite{Geroch-asymp}).\footnote{The peeling theorem requires that the unphysical metric is atleast \(C^3\) at \(\scri\). In our analysis we do not require the full strength of the peeling theorem --- in particular the limits defining \(\mc J_a\) and \(\mc I_{ab}\) in \cref{eq:non-peeled} need not exist on \(\scri\).\label{fn:non-peeled}} In any choice of a foliation of \(\scri\) we define the fields
\begin{subequations}\label{eq:weyl-defn}\begin{align}
    \mc R_{ab} & \defn (\Omega^{-1}C_{cdef}) q_a{}^c n^d q_b{}^e n^f \eqsp & \mc S_a & \defn (\Omega^{-1}C_{cdef}) l^c n^d q_a{}^e n^f \\
    \mc P & \defn (\Omega^{-1}C_{cdef}) l^c n^d l^e n^f \eqsp & \mc P^* & \defn \tfrac{1}{2} (\Omega^{-1}C_{cdef}) l^c n^d \varepsilon^{ef}\\
    \mc J_a & \defn (\Omega^{-1}C_{cdef}) n^c l^d q_a{}^e l^f \eqsp & \mc I_{ab} & \defn (\Omega^{-1}C_{cdef}) q_a{}^c l^d q_b{}^e l^f \label{eq:non-peeled}
\end{align}\end{subequations}
All the above tensors are orthogonal to both \(n^a\) and \(l^a\) in all indices and so can be considered as tensor fields on the cross-sections \(S\) of the chosen foliation of \(\scri\). Further, \(\mc R_{ab}\) and \(\mc I_{ab}\) are symmetric and traceless with respect to the metric \(q^{ab}\) on the cross-sections. These fields have the following conformal weights
\be
    (\mc R_{ab}, \mc I_{ab}) : w = -1 \eqsp (\mc S_a, \mc J_a) : w = -2 \eqsp (\mc P, \mc P^*) : w = -3
\ee
The relation of these tensors to the Weyl scalar components in the Newman-Penrose notation is given in \cref{eq:weyl-GHP}.

For the fields defined in \cref{eq:weyl-defn}, \cref{eq:curl-weyl} implies the following evolution equations along \(\scri\), which can be verified to be conformally-invariant (\cref{eq:weyl-evol-GHP} in the GHP formalism) 
\begin{subequations}\label{eq:weyl-evol}\begin{align}
    (\Lie_n + 2\Phi) \mc S_a & = (\ms D^b + \tau^b) \mc R_{ab} \\
     (\Lie_n + 3\Phi) \mc P~ & = (\ms D^a + 2 \tau^a) \mc S_a - \sigma^{ab} \mc R_{ab} \\
     (\Lie_n + 3\Phi) \mc P^* & = - \varepsilon^{ab}(\ms D_a + 2 \tau_a) \mc S_b + \varepsilon_b{}^c \sigma^{ab} \mc R_{ac} \\
    (\Lie_n + 2 \Phi) \mc J_a & = \tfrac{1}{2}(\ms D_b + 3\tau_b ) (q_a{}^b \mc P - \varepsilon_a{}^b \mc P^*) - 2 \sigma_a{}^b \mc S_b \\
    (\Lie_n + \Phi) \mc I_{ab} & =   (q_a{}^c q_b{}^d - \tfrac{1}{2} q_{ab} q^{cd} )  (\ms D_c + 4 \tau_c) \mc J_d - \tfrac{3}{2} \sigma_{ac} (q_b{}^c \mc P - \varepsilon_b{}^c \mc P^*)
\end{align}\end{subequations}

We define the \emph{News tensor} by
\be\label{eq:News-defn}
    N_{ab} \defn 2 (\Lie_n - \Phi ) \sigma_{ab}
\ee
which satisfies \(N_{ab}n^b = 0\), \(N_{ab}q^{ab} = 0\) and is conformally-invariant on \(\scri\). The News tensor is related to the curvature tensor \(S_{ab}\) (\cref{eq:S-defn}) by (see \cref{eq:News-Ric-GHP} in the GHP formalism)
\be\label{eq:News-Ric}
    N_{ab} =  (q_a{}^c q_b{}^d - \tfrac{1}{2} q_{ab} q^{cd} ) \lb[ S_{cd} - 2\Phi \sigma_{cd} + 2 (\ms D_c \tau_d + \tau_c \tau_d) \rb]
\ee
and to the Weyl tensor on \(\scri\) by (from \cref{eq:Weyl-S})
\be\label{eq:weyl-News}
    \mc R_{ab} = \tfrac{1}{2} \Lie_n N_{ab} \eqsp \mc S_a = \tfrac{1}{2} \ms D^b N_{ab}
\ee

The form of the asymptotic BMS-supertranslation in a general conformal frame can be derived either by transforming between the usual form in the Bondi frame (see \cref{sec:non-Bondi}) or directly from the universal structure at \(\scri\) (\cref{rem:univ-strc}). The final result is as follows: the BMS-supertranslation algebra \(\mf s^\pm\) at \(\scri^\pm\), respectively, is generated by vector fields on the form \(\xi^a = f n^a\) on \(\scri^\pm\), where the function \(f\) satisfies
\be\label{eq:st-null}
    \mf s^\pm = \set{ f \in C^\infty_{w=1}(\scri^\pm) \st (\Lie_n - \Phi) f\vert_{\scri^\pm} = 0 ,~ f\vert_{i^0} = 0 }
\ee
Note that \(f\) has conformal weight \(w =1\), indicated by the subscript in \cref{eq:st-null}. In any fixed choice of conformal factor, such functions \(f\) are completely determined by their value at some chosen cross-section \(S \cong \bb S^2\) of \(\scri\). But, since we are allowed to change the conformal factor on \(\scri\) away from the chosen cross-section \(S\), we cannot (yet) characterise \(\mf s^\pm\) by conformally-weighted functions on \(\bb S^2\), as is usually done in the Bondi conformal frame. We will show later (see \cref{eq:st-null-lim}) how such a characterisation can be obtained on limiting cross-sections of \(\scri\) near \(i^0\).

The BMS-supertranslations contain a (unique; see \cite{Sachs2}) \(4\)-dimensional subalgebra \(\mf t^\pm\) of BMS-translations given by
\be\label{eq:trans-null}
    \mf t^\pm = \set{ f \in \mf s^\pm \st  (q_a{}^c q_b{}^d - \tfrac{1}{2} q_{ab} q^{cd} ) (\ms D_c + \tau_c) (\ms D_d - \tau_d ) f = 0  }
\ee

On some cross-section \(S\) of \(\scri\) the BMS-supermomentum corresponding to some \(f \in \mf s^\pm\) is given by
\be\label{eq:charge-null}
    \mc Q[f; S] = - \int_S \varepsilon_2~ f \lb[ \mc P + \tfrac{1}{2} \sigma^{ab} N_{ab} \rb]
\ee
where \(\df\varepsilon_2 \equiv \varepsilon_{ab}\) is the area element on \(S\) (\cref{eq:null-fields}). The flux of these charges in the region \(\Delta\scri\) of \(\scri\) bounded by two cross-sections \(S_2\) and \(S_1\) is given by the exterior derivative of the integrand in \cref{eq:charge-null} which, using \cref{eq:null-fields,eq:st-null}, can be expressed as \(\varepsilon_3 ~ f (\Lie_n + 3 \Phi) \lb[ \mc P + \tfrac{1}{2} q^{ac} q^{bd} \sigma_{ab} N_{cd} \rb]\) where \(\df\varepsilon_3 \equiv \varepsilon_{abc}\) is the volume element on \(\scri\) (\cref{eq:null-fields}). Then using \cref{eq:Lie-n-q,eq:weyl-evol,eq:weyl-News} we get
\begin{subequations}\label{eq:flux-null-all}\begin{align}
    \mc F[f; \Delta\scri] & \defn \mc Q[f; S_2] - \mc Q[f; S_1] \nonumber \\
    & ~= - \int_{\Delta \scri} \varepsilon_3 ~ f \lb[ \tfrac{1}{4} N^{ab} N_{ab} +  (\ms D^a + 2\tau^a) \mc S_a \rb] \label{eq:flux-null1} \\
    & ~= - \int_{\Delta\scri} \varepsilon_3~ \lb[ \tfrac{1}{4}f N^{ab} N_{ab} - \mc S^a (\ms D_a - \tau_a) f \rb] \label{eq:flux-null} \\
    & ~= - \int_{\Delta\scri} \varepsilon_3~ \lb[ \tfrac{1}{4}f N^{ab} N_{ab} - \tfrac{1}{2} \ms D_b N^{ab} (\ms D_a - \tau_a) f  \rb] \label{eq:flux-null2} \\
    & ~= - \int_{\Delta\scri} \varepsilon_3~ \lb[ \tfrac{1}{4}f N^{ab} N_{ab} + \tfrac{1}{2} N^{ab} (\ms D_a + \tau_a) (\ms D_b - \tau_b) f \rb] \label{eq:flux-null3}
\end{align}\end{subequations}
where \cref{eq:flux-null,eq:flux-null2,eq:flux-null3} are obtained by integrating-by-parts on the cross-sections of \(\scri\) using \cref{eq:IBP,eq:weyl-News}. Note, due to our orientation conventions (\cref{tab:orientation}), if \(S_2\) is in the future of \(S_1\) the fluxes \cref{eq:flux-null-all} measure the \emph{incoming flux} on both \(\scri^\pm\). Further, from \cref{eq:trans-null,eq:flux-null3}, we reproduce the Bondi energy flux for BMS-translations \(f \in \mf t^\pm\).

\begin{remark}[Relation to previously obtained formulae]\label{rem:prev-rel}
We note that in the Bondi-Sachs conformal frame and with the choice of foliation corresponding to constant Bondi time \(\Phi\) and \(\tau_a\) vanish on \(\scri\) (see \S~9.8 \cite{PR2}) and our expressions \cref{eq:News-defn,eq:News-Ric,eq:st-null,eq:trans-null,eq:charge-null,eq:flux-null3} are equivalent to the usual formulae \cite{AS-symp,WZ,AK}. In general conformal choices, if we choose the foliation of \(\scri\) (and hence the auxilliarly normal \(l^a\)) such that \(\tau_a = 0\), then \cref{eq:News-Ric} reduces to the News tensor used in \cite{Ash-Mag-Ash}. Further consider the BMS-momentum in \cref{eq:charge-null}, with the News tensor given by \cref{eq:News-Ric}; using \cref{eq:trans-null} for BMS-translations we can replace the term \((\ms D_a \tau_b + \tau_a \tau_b) f\) by \(\ms D_a \ms D_b f\), reproducing the expression for the BMS-momentum used in Eq.~1 \cite{Ash-Mag-Ash}. 
\end{remark}

\hr

Since the metric \(g_{ab}\) is \(C^{>0}\), \(\Omh C_{abcd}\) is \(C^{>-1}\), at \(i^0\) along spatial directions, and let \(\dd C_{abcd}(\vec\eta) \defn \lim_{\to i^0}\Omh C_{abcd} \) \cite{AH,Ash-in-Held}. The \emph{electric} part of \(\dd C_{abcd}(\vec\eta)\) defined by
\be\label{eq:E-defn}
    \dd E_{ab} (\vec\eta) \defn \dd C_{acbd} (\vec\eta) \dd\eta^c \dd\eta^d 
\ee
is orthogonal to \(\dd\eta^a\) and thus induces an intrinsic field \(\dd E_{ab}\) on \(\hyp\). Since the Weyl tensor is trace-free and satisfies \cref{eq:curl-weyl}, \(\dd E_{ab}\) on \(\hyp\) satisfies (see \cite{AH} for details)
\be\label{eq:EE-hyp}
    \dd h^{ab} \dd E_{ab} = 0 \eqsp \dd D_{[a} \dd E_{b]c} = 0
\ee
We defer the analysis of solutions to these equations to \cref{sec:wave-hyp}.

At spatial infinity Spi-supertranslations are generated by vector fields \(\xi^a\) in \(M\) such that \(\Omega^{-1} \xi^a\) is \(C^{>-1}\) at \(i^0\) in spatial directions and satisfies (see \cref{eq:st-BS} for the corresponding transformations in the coordinates used by Beig and Schmidt \cite{Beig-Schmidt})
\be\label{eq:st-vec-hyp}
    \lim_{\to i^0} \Omega^{-1} \xi^a = \dd f \dd \eta^a - \dd D^a \dd f
\ee
for some smooth function \(\dd f\) on \(\hyp\). Thus, the Spi-supertranslation algebra \(\mf s^0\) is given by all smooth functions \(\dd f\) on \(\hyp\) i.e.
\be\label{eq:symm-hyp}
    \mf s^0 \cong C^\infty(\hyp)
\ee
The \(4\)-dimensional Spi-translation subalgebra \(\mf t^0\) is given by
\be\label{eq:trans-hyp}
    \mf t^0 = \set{ \dd f \in \mf s^0 \st \dd D_a \dd D_b \dd f + \dd h_{ab} \dd f = 0 }
\ee

\begin{remark}[Translation vectors at \(i^0\)]\label{rem:trans-vectors}
Given any element \(\dd f \in \mf t^0\), consider the vector at \(i^0\) defined by (note the sign difference in the hyperboloidal component relative to \cref{eq:st-vec-hyp})
\be\label{eq:trans-vector-defn}
    \dd v^a \defn \dd f \dd \eta^a + \dd D^a \dd f
\ee
Using \cref{eq:d-eta-h} we have
\be
    \dd\partial_a \dd v_b = \dd D_a \dd D_b \dd f + \dd h_{ab} \dd f = 0
\ee
so, \(\dd v^a\) is \emph{direction-independent} i.e., \(\dd v^a \in Ti^0\).
Thus, the Spi-translations \(\mf t^0\) can be represented by vectors in \(Ti^0\).
\end{remark}

The conserved charges associated to Spi-translations \(\dd f \in \mf t^0\) were given by Ashtekar and Hansen \cite{AH}. We define the Spi-supermomenta for any Spi-supertranslation \(\dd f \in \mf s^0\) using the same expression as \cite{AH} --- in \cref{sec:CD} we show that our final result is unchanged if we use, instead, the expression for the Spi-supermomenta given by Comp\`ere and Dehouck \cite{CD}. Thus, on a cross-section \(S\) of \(\hyp\), with unit future-pointing timelike normal \(\dd u^a\), we take the Spi-supermomenta to be given by 
\be\label{eq:charge-hyp}
	\mc Q[\dd f; S] = \int_S \df{\dd \varepsilon}_2~   \dd E_{ab} \dd u^a \dd D^b \dd f
\ee
where \(\df{\dd\varepsilon}_2 \equiv \dd\varepsilon_{ab}\) is the area element on \(S\) (\cref{eq:volume-hyp}). From \cref{eq:EE-hyp} it follows that \(\dd D^a \dd E_{ab} = 0\), and the flux of the Spi-supermomenta between any region \(\Delta\hyp\) of \(\hyp\) bounded by the cross-sections \(S_2\) and \(S_1\) (where \(S_2\) is in the future of \(S_1\)) is
\be\label{eq:flux-hyp}
    \mc F[\dd f; \Delta\hyp] \defn \mc Q[\dd f; S_2] - \mc Q[\dd f; S_1] = \int_{\Delta\hyp} \df{\dd\varepsilon}_3~  \dd E_{ab} \dd D^a \dd D^b \dd f
\ee
where \(\df{\dd\varepsilon}_3 \equiv \dd\varepsilon_{abc}\) is the volume element on \(\hyp\). 

For Spi-translations \(\dd f\in \mf t^0\), using \cref{eq:EE-hyp,eq:trans-hyp}, the fluxes \cref{eq:flux-hyp} vanish across \emph{any} region \(\Delta\hyp\). Thus corresponding Spi-momenta \cref{eq:charge-hyp} are independent of the choice of cross-section of \(\hyp\) and are well-defined at \(i^0\); these are the usual ADM-momenta at \(i^0\)  \cite{AH,AMA-spi-3+1}. For any Spi-translation \(\dd f\) let \(\dd v^a\) be the direction-independent vector at \(i^0\) determined by \cref{eq:trans-vector-defn}. The charge \cref{eq:charge-hyp} is then a linear map from \(Ti^0\) to the reals which determines the \emph{direction-independent} ADM-momentum covector \(\dd P_a\) at \(i^0\) by
\be\label{eq:P-vector}
    \dd P_a \dd v^a = \mc Q[\dd f; S]
\ee
where \(S\) is any cross-section of \(\hyp\). The Spi-supermomenta defined by \cref{eq:charge-hyp} where \(\dd f\) is a general Spi-supertranslation can only be associated to the blowup \(\hyp\) and not to the asymptotic boundary \(i^0\) itself; however, these will be useful when relating the BMS-supermomenta on \(\scri^-\) to those on \(\scri^+\).\\

\begin{remark}[Supertranslation vector fields near \(i^0\)]
\label{rem:st-vec}

Let \(f\) be any function in the unphysical spacetime \(M\) with conformal weight \(w=1\), which satisfies \cref{eq:st-null} on \(\scri\), and is \(C^{>0}\) in spatial directions at \(i^0\). Consider the vector field in \(M\) given by
\be\label{eq:st-vec-M}
    \xi^a \defn f n^a - \Omega \nabla^a f
\ee
Clearly \(\xi^a\vert_{\scri}\) generates a BMS-supertranslation on \(\scri\). Since, \(f\) is \(C^{>0}\) along spatial directions at \(i^0\) and \(f\vert_{i^0} = 0\), \(\Omega^{-\half} f\) is \(C^{>-1}\) at \(i^0\). Then, defining \(\dd f(\vec\eta) \defn \lim_{\to i^0} \Omega^{-\half} f \) along spatial directions, using \cref{eq:eta-defn,eq:dd-der-spatial}, we can verify \cref{eq:st-vec-hyp}, that is, \(\xi^a\) generates a Spi-supertranslation along spatial directions at \(i^0\). Further, since under a change of conformal factor \(\Omega \mapsto \omega \Omega\) we have \(n^a \mapsto \omega^{-1} (n^a + \Omega \nabla^a \ln \omega) \) and \(f\) has conformal weight \(w=1\), \(\xi^a\) is independent of the choice of conformal factor used in the conformal-completion of the physical spacetime. Thus, asymptotically the vector field defined by \cref{eq:st-vec-M} generates a BMS-supertranslation on \(\scri\) and a Spi-supertranslation at \(i^0\) along spatial directions.
\end{remark}

Thus, in this picture the asymptotic supertranslation algebra including all the asymptotic regions is the direct sum
\be\label{eq:symm-total}
    \mf s = \mf s^- \oplus \mf s^0 \oplus \mf s^+ \cong C^\infty_{w=1}(\scri^-) \oplus C^\infty(\hyp) \oplus C^\infty_{w=1}(\scri^+)
\ee
This structure of \(\mf s\) arises essentially because the hyperboloid \(\hyp\) does not ``attach'' to null directions along \(\scri\) at \(i^0\) and one cannot demand any ``continuity'' between the supertranslations as they approach \(i^0\) along null directions and spatial directions. An additional issue not present in the case of Maxwell fields studied in \cite{KP-EM-match}, is the presence of the conformal weights for the BMS-supertranslations \(\mf s^\pm\) which is absent for the Spi-supertranslations \(\mf s^0\). Both of these issues can be resolved by using the space \(\cyl\) (see \cref{sec:cyl}) which includes both null and spatial directions at \(i^0\) as follows.

Consider first a BMS-supertranslation \(f^\pm \in \mf s^\pm\) on \(\scri^\pm\). From \cref{eq:st-null}, \(f^\pm\vert_{i^0} = 0\) and so let \(F^\pm = \Sigma f^\pm\). Using \cref{eq:st-null,cond:Sigma-choice} we have on \(\scri\)
\be\label{eq:Lie-n-F}
    \Lie_n F^\pm = (\Phi - 2) F^\pm
\ee
Integrating this along the null generators, since \(\Phi\vert_{i^0} = 2\), we have \(\dd F^\pm = \lim_{\to i^0} F^\pm\) along \(\scri^\pm\) is a smooth function on \(\nulls^\pm\). The limiting rescaled functions \(\dd F^\pm\) are invariant under a change of conformal factor (since \(\omega\vert_{i^0} = 1\)) but transform as \(\dd F^\pm \mapsto \dd\sigma^\pm \dd F^\pm\) under a change of rescaling function \(\Sigma \mapsto \sigma \Sigma\), that is, \(\dd F^\pm\) has rescaling weight \(r=1\) on \(\nulls^\pm\). Thus, the BMS-supertranslations can be characterised near \(i^0\) as
\be\label{eq:st-null-lim}
    \mf s^\pm = \set{ \dd F^\pm \defn \lim_{\to i^0} \Sigma f^\pm }  \cong C^\infty_{r=1} (\nulls^\pm)
\ee
Since the metric \(\tilde{\dd q}_{ab}\) on \(\nulls^\pm\) transforms according to \(\tilde{\dd q}_{ab} \mapsto \dd \sigma^\pm \tilde{\dd q}_{ab}\) (see \cref{eq:rescaled-q}), in the limit to \(i^0\) we recover the characterisation of BMS-supertranslations as conformally-weighted functions on \(\nulls^\pm \cong \bb S^2\) where now the rescaling freedom \(\dd \sigma^\pm\) is viewed as a conformal transformation on \(\nulls^\pm\). We show in \cref{sec:BS} that the functions \(\dd F^\pm\) are precisely the usual BMS-supertranslation functions used in the Bondi conformal choice.

Next consider a Spi-supertranslation \(\dd f \in \mf s^0\) which is a smooth function on \(\hyp\) and hence on \(\cyl\setminus\nulls^\pm\). However, \(\dd f\) does not have a limit to \(\nulls^\pm\) even for Spi-translations, though \(\dd\Sigma \dd f\) does extend smoothly to \(\nulls^\pm\) (see \cref{eq:trans-hyp-coord,eq:F-null-coord}). Thus, we restrict to Spi-supertranslations \(\dd f\) such that the rescaled function \(\dd F \defn \dd\Sigma \dd f\) extends smoothly to \(\nulls^\pm\).\footnote{This condition on the Spi-supertranslations at \(\nulls^\pm\) is also suggested in the last footnote of \cite{Ash-in-Held}.} As in the case of a BMS-supertranslation the rescaled function \(\dd F\) is invariant under changes of the conformal factor but has rescaling weight \(r=1\). Note that such Spi-supertranslation \(\dd F \in \mf s^0\) determines unique BMS-supertranslations \(\dd F^\pm \in \mf s^\pm\) by ``continuity'' at \(\nulls^\pm\) that is, \(\dd F\vert_{\nulls^\pm} = \dd F^\pm\). Thus, there is a natural subalgebra \(\mf s^{\rsub{nr}}\) of \(\mf s\) (\cref{eq:symm-total}) given by the \emph{null-regular} supertranslations of the form
\be\label{eq:st-null-regular}
    \mf s^{\rsub{nr}} = \set{ (\dd F^-, \dd F, \dd F^+) \in \mf s  \st \dd F^\pm = \dd F\vert_{\nulls^\pm} } \cong C^\infty_{r=1}(\cyl)  
\ee
Note that we have replaced the conformal weight \(w=1\) of the BMS-supertranslations represented by \(f^\pm\) with the rescaling weight \(r=1\) of the limiting functions \(\dd F^\pm\), and similarly introduced the Spi-supertranslation function \(\dd F\) of rescaling weight \(r=1\). Thus, the null-regular subalgebra in \cref{eq:st-null-regular} is both conformal and rescaling invariant.

As shown in \cite{Ash-Mag-Ash,AS-ang-mom}, there exists an isomorphism between Spi-translations \(\mf t^0\) and both the BMS-translations \(\mf t^\pm\). We illustrate this isomorphism explicitly using \cref{eq:st-null-regular} as follows. Since \cref{eq:st-null-regular} is invariant under a change of rescaling function, we choose \(\Sigma\) such that the metric \(\tilde{\dd q}_{ab}\) on \(\nulls^\pm\) is that of a unit \(2\)-sphere as in \cref{sec:wave-hyp}. Then as shown in \cref{eq:F-null-coord}, for any element of the Spi-translations \(\mf t^0\), the functions \(\dd F\vert_{\nulls^\pm}\) are spanned by the \(\ell = 0,1\) spherical harmonics. For such functions the BMS-supertranslations \(f^\pm\) such that \(\lim_{\to i^0} \Sigma f^\pm = \dd F^\pm = \dd F\vert_{\nulls^\pm}\) are precisely the ones which satisfy \cref{eq:trans-null} (note that \(\tau_a = 0\) in our choice of foliation; \cref{eq:tau-0}). Thus, any Spi-translation in \(\mf t^0\) determines \emph{unique} elements of both \(\mf t^\pm\) satisfying \cref{eq:st-null-regular}. Thus, \(\mf s^{\rsub{nr}}\) contains a unique subalgebra of translations which is isomorphic to \(\mf t^0\) and \(\mf t^\pm\). Note from \cref{eq:F-null-coord} this isomorphism naturally provides an antipodal identification between \(\mf t^-\) and \(\mf t^+\).

For supertranslations, which are not translations, \cref{eq:st-null-regular} does not provide a unique isomorphism from \(\mf s^-\) to \(\mf s^+\). Given a BMS-supertranslation \(\dd F^-\) on \(\scri^-\) we can get any BMS-supertranslation \(\dd F^+\) on \(\scri^+\) simply by choosing \(\dd F\) suitably on \(\cyl\) since general Spi-supertranslations can be arbitrary functions on \(\cyl\). Since the conformal reflection isometry of \(\cyl\) provides a natural map \(\Upsilon : \nulls^- \to \nulls^+\) between the space of null generators of \(\scri^-\) and \(\scri^+\), one can impose some matching condition relating \(\Upsilon \circ \dd F^-\) to \(\dd F^+\). However, our considerations so far do not provide any physical criteria for such a matching condition --- except by analogy to translations. As we show below, using the Einstein equation on \(\cyl\) and demanding that the total flux of Spi-supermomenta through \(\cyl\) vanish in any scattering process provides a natural matching condition (\cref{eq:symm-antipodal}) which is precisely the one conjectured in \cite{Stro-CK-match}.

\section{Null-regular spacetimes at \(i^0\)}
\label{sec:matching}

Even with the ``partial matching'' of the supertranslations provided by \cref{eq:st-null-regular}, we cannot immediately conclude that the corresponding supermomenta match in the same manner at \(i^0\). In fact, since we have not specified the behaviour of the gravitational fields on \(\scri\) in the limit to \(i^0\), the BMS-supermomenta \cref{eq:charge-null} can diverge as the cross-sections of \(\scri\) tend towards \(i^0\). Such spacetimes will have an infinite flux of BMS-supermomenta through null infinity and are thus unphysical in a scattering process. We will now impose the following additional restrictions on the class of spacetimes we consider, from here on, to discard such ``pathological'' solutions to the Einstein equation. Even though we argue that these conditions are physically reasonable for a scattering process, the existence of (a suitably large class of) such spacetimes is very much an open question which we discuss in \cref{sec:disc}.

\begin{definition}[Null-regular spacetime at \(i^0\)]
\label{def:null-regular}
Let \(l^a\) be the vector field defined by \cref{eq:l-defn} in \(M\). We call a spacetime with an Ashtekar-Hansen structure (\cref{def:AH}) \emph{null-regular} at \(i^0\) if
\begin{enumerate}
    \item the rescaled quantity
    \be\label{eq:P-falloff}
    \Sigma^{-3} \Omega^{-1} C_{abcd}l^a n^b l^c n^d \text{ is } C^{>-1} \text{ in \emph{both} null and spatial directions at } i^0
    \ee
    \item in the limit to \(i^0\) along each null generator of \(\scri\), the components of \(\Sigma^{1+\epsilon} N_{ab}\) and \(\Sigma^{1+\epsilon} \mc R_{ab}\) in a \(C^{>1}\)-chart at \(i^0\) remain bounded for some small \(\epsilon > 0\), that is,\footnote{Note that the falloffs of \(N_{ab}\) and \(\mc R_{ab}\) are compatible with \cref{eq:weyl-News}. However, integrating \cref{eq:weyl-News} to derive the falloffs on the News tensor \(N_{ab}\) from those on \(\mc R_{ab}\), in general, leads to a ``constant of integration'' in \(N_{ab}\). The faster falloff of the News tensor imposed in \cref{eq:N-R-falloff} ensures that this constant of integration vanishes.}
    \be\label{eq:N-R-falloff}
     N_{ab} = O(\Sigma^{- (1+\epsilon)}) \eqsp \mc R_{ab} = O(\Sigma^{-(1+\epsilon)}) \text{ as } \Sigma^{-1} \to 0 \text{ along } \scri
    \ee
\end{enumerate}
\end{definition}

Our falloff condition on the News tensor is stronger than the one required for a finite flux of Bondi momentum (associated to translations) through \(\scri\) \cite{Ash-Mag-Ash}, but is weaker than the falloff assumed in \cite{Stro-CK-match} which was motivated by CK-spacetimes \cite{CK}. As we show below (\cref{rem:finite-flux-scri}), this falloff is needed so that the flux of \emph{all} BMS-supermomenta through \(\scri\) is finite. The falloffs in \cref{eq:N-R-falloff} also ensure that the radiative phase space on \(\scri\) has a well-defined symplectic structure \cite{AS-symp,ACL}. The form of these conditions in terms of the Bondi-Sachs parameter on \(\scri\) is given in \cref{eq:null-regular-BS}. \\

Let us explore the consequences of the regularity conditions \cref{def:null-regular} on the gravitational fields on \(\scri\) and at \(i^0\). From \cref{eq:weyl-defn} we have on \(\scri\), \(\Sigma^{-3} \Omega^{-1} C_{abcd}l^a n^b l^c n^d = \Sigma^{-3} \mc P\), and from \cref{eq:P-falloff} \(\lim_{\to i^0} \Sigma^{-3} \mc P\) along \(\scri\) exists and induces a smooth function on the space of null directions \(\nulls^\pm\). Similarly, the limit of \(\Sigma^{-3} \Omega^{-1} C_{abcd} l^a n^b l^c n^d\) to \(i^0\) in spatial directions can be rewritten as (using \cref{eq:dd-Sigma,eq:N-defn,eq:l-defn,eq:E-defn,eq:U-defn})
\be
    \lim_{\to i^0} \Sigma^{-3}\Omega^{-1} C_{abcd} l^a n^b l^c n^d = \lim_{\to i^0} \Sigma^{-3} \Omega^{-1} C_{abcd} L^a N^b L^c N^d =  (\dd \Sigma^{-1} \dd E_{ab} \dd U^a \dd U^b) (\vec\eta)
\ee
Thus, \(\Sigma^{-3}\Omega^{-1} C_{abcd} l^a n^b l^c n^d\) induces the field \(\dd \Sigma^{-1} \dd E_{ab} \dd U^a \dd U^b\) on \(\hyp\) which can be viewed as a field on \(\cyl \setminus \nulls^\pm\). Since \(\Sigma^{-3}\Omega^{-1} C_{abcd} l^a n^b l^c n^d\) is \(C^{>-1}\) in both null and spatial directions (\cref{eq:P-falloff}), \(\dd \Sigma^{-1} \dd E_{ab} \dd U^a \dd U^b\) has a limit as a smooth function on \(\nulls^\pm\), which further coincides with the field induced by \(\lim_{\to i^0}\Sigma^{-3}\Omega^{-1} C_{abcd} l^a n^b l^c n^d = \lim_{\to i^0} \Sigma^{-3} \mc P\) on \(\nulls^\pm\) from \(\scri^\pm\) that is,
\be\label{eq:E-field-match}
    \lim_{\to i^0} (\Sigma^{-3} \mc P) \text{ along } \scri^\pm = \lim_{\to \nulls^\pm}(\dd \Sigma^{-1} \dd E_{ab} \dd U^a \dd U^b) \text{ along } \cyl
\ee

Now, consider the foliation of \(\scri\) by cross-sections \(S_\Sigma\) with \(\Sigma^{-1} = \text{constant}\) as described \cref{sec:cyl}. From \cref{eq:News-defn,cond:Sigma-choice} it follows that the shear tensor \(\sigma_{ab}\) of such cross-sections satisfies
\be
    \Lie_n (\Sigma \sigma_{ab}) = \tfrac{1}{2}\Sigma N_{ab} +  (\Phi - 2 ) \Sigma \sigma_{ab}
\ee
We can integrate this equation along the null generators (using \cref{eq:N-n-Sigma,eq:N-R-falloff}), noting that \(\Phi\vert_{i^0} = 2\), to conclude that
\be\label{eq:sigma-falloff}
    \Sigma \sigma_{ab} \text{ remains bounded as } \Sigma^{-1} \to 0
\ee\\

Using these conditions we can evaluate the supermomenta induced on the space of null directions \(\nulls^\pm\) in the limit from \(\scri^\pm\) and from \(\cyl\). First, consider the BMS-supermomentum associated to a BMS-supertranslation \(f^\pm \in \mf s^\pm\) on the cross-sections \(S_\Sigma\) of \(\scri^\pm\) tend towards \(i^0\), i.e., as \(\Sigma^{-1} \to 0\). From \cref{eq:charge-null}, the BMS-supermomentum induced on \(\nulls^\pm\) from \(\scri^\pm\) is given by
\be\label{eq:charge-null-scri}\begin{split}
    \mc Q[f^\pm; \nulls^\pm] 
    & = - \lim_{\Sigma^{-1} \to 0}~ \int_{S_\Sigma} \varepsilon_2~ f^\pm \lb[ \mc P + \tfrac{1}{2} q^{ac} q^{bd} \sigma_{ab} N_{cd} \rb] \\
    & = - \lim_{\Sigma^{-1} \to 0}~ \int_{S_\Sigma} (\Sigma^2\df\varepsilon_2)~ (\Sigma f^\pm) (\Sigma^{-3} \mc P + \tfrac{1}{2} \Sigma \tilde q^{ac} \tilde q^{bd} \sigma_{ab} N_{cd} ) \\
    & = - \int_{\nulls^\pm} \tilde{\df{\dd \varepsilon}}_2~ \dd F^\pm (\Sigma^{-3} \mc P ) 
\end{split}\ee
where in the third line we have used the falloffs \cref{eq:N-R-falloff,eq:sigma-falloff}, the rescaled area element \(\tilde{\df{\dd\varepsilon}}_{ab} = \lim_{\to i^0} \Sigma^2 \df\varepsilon_{ab}\) and the rescaled supertranslation function \(\dd F^\pm = \lim_{\to i^0} \Sigma f^\pm \) on \(\nulls^\pm\). The final limit exists due to our null-regularity condition \cref{eq:P-falloff} as discussed above.

\begin{remark}[Finiteness of flux through \(\scri\)]
\label{rem:finite-flux-scri}
Let \(\Delta \scri^\pm\) be a region of \(\scri^\pm\) foliated by cross-sections \(S_\Sigma\) with \(\Sigma^{-1}_0 \leq \Sigma^{-1} \leq \Sigma^{-1}_1\). Since the limit to \(i^0\) of the BMS-supermomenta evaluated on \(S_\Sigma\) exists, the flux must be finite as \(\Sigma^{-1}_0 \to 0\). However, the volume element \(\varepsilon_{abc}\) on \(\scri^\pm\) (appearing in \cref{eq:flux-null}) is ill-defined in this limit since \(l^a\) diverges at \(i^0\). But the rescaled volume element
\be
    L^d \varepsilon_{dabc} = 2 \Sigma^{-1} \varepsilon_{abc} = 3 \nabla_{[a} \Sigma^{-1} \varepsilon_{bc]} \text{ is } C^{>-1} \text{ at } i^0
\ee
Thus, on \(\Delta \scri^\pm\) foliated by \(S_\Sigma\) we can rewrite the flux \cref{eq:flux-null}, using the rescaled quantities \(F^\pm = \Sigma f^\pm\), \(\tilde q^{ab} = \Sigma^{-2} q^{ab}\) and \(\tilde\varepsilon_{ab} = \Sigma^2 \varepsilon_{ab}\), as 
\be\begin{split}
    \mc F[f^\pm; \Delta\scri^\pm] 
    & = - \tfrac{1}{2} \int_{\Sigma^{-1}_0}^{\Sigma^{-1}_1} d \Sigma^{-1}~ \Sigma \int_{S_\Sigma} \tilde{\df\varepsilon}_2~ \lb[ \tfrac{1}{4} F^\pm \Sigma ~ \tilde q^{ac} \tilde q^{bd} N_{ab} N_{cd} - \Sigma^{-1} \tilde q^{ab} \mc S_a \nabla_b F^\pm \rb]
\end{split}\ee
Since the flux on the left-hand-side is finite, in the limit \(\Sigma^{-1}_0 \to 0\), and \(F^\pm\), \(\tilde q^{ab}\) and \(\tilde{\df\varepsilon}_2\) on the right-hand-side induce smooth fields on \(\nulls^\pm\) and from the falloff on \(N_{ab}\) \cref{eq:N-R-falloff}, we get
\be\label{eq:S-falloff}
    \Sigma^{-1} \mc S_a = O(\Sigma^{-\epsilon}) \quad\text{along null directions}
\ee
for some small \(\epsilon > 0\). Similarly, using the alternative forms for the flux from \cref{eq:flux-null-all} we can also conclude that in the limit \(\Sigma^{-1} \to 0\) towards \(i^0\) along null directions\footnote{If the unphysical metric is also \(C^{>0}\) in null directions then, \cref{eq:null-regular-extra} would follow from \cref{eq:weyl-News,eq:S-falloff,eq:N-R-falloff}.} 
\be\label{eq:null-regular-extra}\begin{split}
    \Sigma^{-1} \tilde q^{ab} \ms D_a \mc S_b = O(\Sigma^{-\epsilon}) &\eqsp \Sigma \tilde q^{bc} \ms D_b N_{ac} = O(\Sigma^{-\epsilon}) \\
     (\tilde q_a{}^c \tilde q_b{}^d - \tfrac{1}{2} \tilde q_{ab} \tilde q^{cd} ) \ms D_c \ms D_d F^\pm & \quad\text{ remains bounded} 
\end{split}\ee
 Note that, from \cref{eq:trans-null,eq:flux-null-all}, the term linear in the News tensor in the flux of BMS-supermomenta vanishes identically for BMS-translations. Thus, if one is only interested in the BMS-translations, the slower falloff \(N_{ab} = O(\Sigma^{-(1+\epsilon)/2})\) (equivalent to the falloff used in \cite{Ash-Mag-Ash}) suffices to ensure that the BMS-momentum flux is finite.
\end{remark}

\begin{remark}[Classical vacua on \(\scri\)]
\label{rem:vacua}
From \cref{eq:weyl-News}, if the \(N_{ab} = 0\) vanishes in some open region of \(\scri\) then \(\mc R_{ab} = \mc S_a = 0\). Conversely if \(\mc R_{ab} = \mc S_a = 0\) in a region of \(\scri\) which contains the entire space of generators (topologically \(\bb S^2\)), then the News tensor \(N_{ab}\) is a symmetric traceless tensor on \(\bb S^2\) (as \(\Lie_n N_{ab} = 0\)) satisfying \(\ms D^b N_{ab} = 0\). Since the News tensor is conformally-invariant, we can choose the unit-metric on \(\bb S^2\) to show that any such smooth tensor must vanish; see for instance, Appendix~A.4 \cite{AK}, Appendix~C \cite{AMA-spi-3+1}, or Prop.~4.15.59 \cite{PR1}. If an asymptotically-flat spacetime is such that its News tensor vanishes on \emph{all} of \(\scri\), then it corresponds to a \emph{classical vacuum} in the radiative phase space on \(\scri\) (see \cite{AS-symp} or \cite{ACL} for a recent review). Our falloff conditions \cref{eq:N-R-falloff,eq:S-falloff,eq:sigma-falloff} (equivalently \cref{eq:null-regular-BS}) then imply that all null-regular spacetimes approach some classical vacuum near \(i^0\), and the tensor \(\Sigma \sigma_{ab}\) (which remains bounded; \cref{eq:sigma-falloff}) measures the deviation of the given spacetime relative to this vacuum.
\end{remark}

Next we evaluate the Spi-supermomentum associated to a Spi-supertranslation \(\dd f \in \mf s^0\) on \(\cyl\) as the cross-section \(S \to \nulls^\pm\). Since, the normals \(\dd u^a\) and \(\dd U^a\) are timelike and unit with respect to the metrics \(\dd h_{ab}\) and \(\tilde{\dd h}_{ab} = \dd\Sigma^2 \dd h_{ab}\), respectively, we have
\be
    \lim_{\to \nulls^\pm} \dd\Sigma^{-1}\dd U^a = \pm \lim_{\to \nulls^\pm} \dd\Sigma^{-2} \dd u^a
\ee
where the signs on the right-hand-side are due to our orientation conventions from \cref{tab:orientation}. Similarly, we have for the area element on \(S\) in the limit to \(\nulls^\pm\) (from \cref{eq:volume-hyp,eq:nulls-area})
\be
    \tilde{\dd\varepsilon}_{ab} = \pm \lim_{S \to \nulls^\pm} \dd\Sigma^2 \dd\varepsilon_{ab} 
\ee
Thus, rewriting the Spi-supermomenta \cref{eq:charge-hyp} in terms of \(\dd F = \dd\Sigma \dd f\), using \cref{eq:U-defn} we have on \(\nulls^\pm\) 
\be\label{eq:charge-null-C}\begin{split}
    \mc Q[\dd f; \nulls^\pm] 
    & = - \int_{\nulls^\pm} \tilde{\df{\dd \varepsilon}}_2~ \lb[ (\dd\Sigma^{-1}\dd E_{ab} \dd U^a \dd U^b) \dd F  \rb. \\
     &\qquad \lb. - \dd\Sigma^2 (\dd\Sigma^{-1} \dd E_{ab} \dd U^a \dd U^b) (\dd\Sigma^{-1}\dd U^c) \dd D_c \dd F  - \dd E_{ab} \dd U^a \tilde{\dd q}^{bc} \dd D_c \dd F \rb] \\
    & = - \int_{\nulls^\pm} \tilde{\df{\dd \varepsilon}}_2~ \dd F (\dd\Sigma^{-1}\dd E_{ab} \dd U^a \dd U^b)
\end{split}\ee
where in the last line we have restricted to the Spi-supertranslations for which \(\dd F\) admits a smooth limit to \(\nulls^\pm\) as discussed at the end of \cref{sec:super-trans}. We have also used \cref{eq:U-on-N} and \cref{eq:E-cyl-falloff} for the electric field on \(\cyl\) in null-regular spacetimes.

Comparing \cref{eq:charge-null-scri,eq:charge-null-C} using \cref{eq:E-field-match}, we see that the supermomenta induced on \(\nulls^\pm\) by the BMS-supermomenta on \(\scri\) and the Spi-supermomenta on \(\cyl\) are equal iff \(\dd F^\pm = \dd F\vert_{\nulls^\pm}\), that is, for the null-regular supertranslations in \cref{eq:st-null-regular} we have
\be\label{eq:charge-partial-matching}\begin{split}
    \mc Q[f^\pm; \nulls^\pm] \text{ along } \scri^\pm & = \mc Q[\dd f; \nulls^\pm] \text{ along } \cyl \quad\text{for } \dd F \in \mf s^{\rsub{nr}}
\end{split}\ee

For a Spi-translation the flux of supermomenta on \(\cyl\) vanishes and so the Spi-momenta on \(\nulls^\pm\) are equal. Thus, from \cref{eq:charge-partial-matching}, the BMS-momenta \cref{eq:charge-null} on \(\scri^\pm\) and the Spi-momenta \cref{eq:charge-hyp} on \(\cyl\) determine the unique ADM-momentum covector \(\dd P_a\) at \(i^0\) through \cref{eq:P-vector} which reproduces the result of \cite{Ash-Mag-Ash}. For a general Spi-supertranslation the Spi-supermomenta on \(\nulls^\pm\) need not be equal. We argue next that the total flux of Spi-supermomenta on \(\cyl\) is ``non-dynamical'' and that the subalgebra of Spi-supertranslations on \(\cyl\) which are \emph{totally fluxless} provides a natural isomorphism from \(\mf s^-\) to \(\mf s^+\).

\subsection{Totally fluxless supertranslations on \(\cyl\), antipodal matching and conservation laws}

As mentioned above, for some arbitrary choice of Spi-supertranslation \(\dd F \in \mf s^{\rsub{nr}}\), the flux of supermomenta through all of \(\cyl\) need not vanish. The flux through all of \(\cyl\) is given by the difference of the Spi-supermomenta integrals on \(\nulls^\pm\)
\be
    \mc F[\dd f; \cyl] = - \int_{\nulls^+} \df{\tilde{\dd \varepsilon}}_2~ \dd F^+ (\dd\Sigma^{-1} \dd E_{ab} \dd U^a \dd U^b) + \int_{\nulls^-} \df{\tilde{\dd \varepsilon}}_2~ \dd F^- (\dd\Sigma^{-1} \dd E_{ab} \dd U^a \dd U^b) 
\ee
where \(\dd F^\pm = \dd F \vert_{\nulls^\pm}\). From the analysis in \cref{sec:wave-hyp}, the only solutions to Einstein equation on \(\cyl\) for which \(\lim_{\to \nulls^\pm}\dd\Sigma^{-1} \dd E_{ab} \dd U^a \dd U^b\) exists (and thus corresponds to null-regular spacetimes) are the ones which satisfy
\be\label{eq:reflection-electric}
    \Upsilon \circ (\dd\Sigma^{-1} \dd E_{ab} \dd U^a \dd U^b)\vert_{\nulls^-} = (\dd\varsigma^{-3} \dd\Sigma^{-1} \dd E_{ab} \dd U^a \dd U^b)\vert_{\nulls^+}
\ee
where \(\Upsilon\) is the reflection conformal isometry of \(\cyl\) (\cref{eq:reflection-cyl}). Then, using the transformation of \(\tilde{\dd\varepsilon}_{ab}\) under \(\Upsilon\), \cref{eq:reflection-cyl-boundary}, the total flux is given by
\be\label{eq:tot-flux-cyl}
    \mc F[\dd f; \cyl] = - \int_{\nulls^+} \df{\tilde{\dd \varepsilon}}_2~ (\dd F^+ + \dd\varsigma^{-1} \Upsilon \circ \dd F^-) (\dd\Sigma^{-1} \dd E_{ab} \dd U^a \dd U^b)
\ee
Since \cref{eq:tot-flux-cyl} is invariant under a change of rescaling function, we consider its value in the choice of \(\Sigma\) from \cref{sec:dd}, where the metric on \(\nulls^\pm\) is that of a unit-sphere and \(\dd\varsigma = 1\). In this choice, for any Schwarzschild spacetime which is ``at rest'' with respect to the chosen \(C^{>1}\)-coordinates at \(i^0\), the scalar \(\dd\Sigma^{-1} \dd E_{ab} \dd U^a \dd U^b\) is a \(\ell=0\) spherical harmonic on \(\nulls^+\) (see \cref{sec:wave-hyp}). However, for general Spi-supertranslations \((\dd F^+ +  \Upsilon \circ \dd F^-)\) can be any arbitrary function on \(\nulls^+\), and the total flux of Spi-supermomenta through \(\cyl\) can take any value even for Schwarzschild spacetime. This suggests that the total flux through \(\cyl\) is ``spurious'' and is not associated to any dynamics of the gravitational scattering process. Thus, we further restrict the symmetry algebra to those elements which have vanishing total flux on \(\cyl\). From \cref{eq:tot-flux-cyl}, the only Spi-supertranslations \(\dd f\) which satisfy \(\mc F[\dd f; \cyl] = 0\) are the ones for which
\be\label{eq:symm-antipodal}
    \Upsilon \circ \dd F^- = - \dd\varsigma \dd F^+ 
\ee
Note that this is \emph{not} a restriction on the spacetimes we consider, unlike the null-regularity conditions in \cref{def:null-regular}. The behaviour of \(\dd F\) in \(\cyl\setminus\nulls^\pm\) can be arbitrary, and only the boundary values at \(\nulls^\pm\) are required to satisfy \cref{eq:symm-antipodal}. Thus, we define the equivalence class \([\dd F]\) for any \(\dd F\) satisfying \cref{eq:symm-antipodal} by
\be\label{eq:equiv-fluxless-symm}
    \dd F' \in [\dd F] \iff \dd F'\vert_{\nulls\pm} = \dd F\vert_{\nulls\pm} 
\ee
Each equivalence class \([\dd F]\) is uniquely determined by a smooth function on \(\bb S^2\) (with rescaling weight \(r=1\)), either considered as a function on \(\nulls^-\) or \(\nulls^+\) related by \cref{eq:symm-antipodal}. Thus, the condition that the total flux through \(\cyl\) vanish gives us the ``diagonal'' subalgebra of the null-regular supertranslation algebra \(\mf s^{\rsub{nr}}\)
\be\label{eq:tot-fluxless-symm}
    \mf s^\times \defn \set{ [\dd F] \st \dd F \in \mf s^{\rsub{nr}} \text{ and } \Upsilon \circ \dd F^- = - \dd\varsigma \dd F^+  }  \cong C^\infty_{r=1}(\bb S^2)
\ee
The supertranslations in \(\mf s^\times\) provide a natural isomorphism between the BMS-supertranslations \(\mf s^-\) and \(\mf s^+\) on null infinity as follows. In any choice of rescaling function \(\Sigma\), any BMS-supertranslation \(f^-\) on \(\scri^-\) determines a unique \([\dd F] \in \mf s^\times\) on \(\cyl\) so that along \(\scri^-\) we have \(\lim_{\to i^0} \Sigma f^- = \dd F \vert_{\nulls^-} = \dd F^-\). From \cref{eq:tot-fluxless-symm} this determines a unique symmetry \(f^+\) on \(\scri^+\) so that along \(\scri^+\) we have \(\lim_{\to i^0} \Sigma f^+ = \dd F\vert_{\nulls^+} = \dd F^+\). Thus we have the isomorphism
\be\label{eq:symm-match}
    \mf s^- \to \mf s^+ : ( \lim_{\to i^0} \Sigma f^-)(\theta^A) \text{ along } \scri^- \mapsto - (\lim_{\to i^0} \Sigma f^+)(-\theta^A) \text{ along } \scri^+
\ee
That is, for the subalgebra \(\mf s^\times\), the symmetries on \(\scri^-\) can be matched to those on \(\scri^+\) through an antipodal reflection on \(\bb S^2\). Note that the negative sign in \cref{eq:symm-match} is due to our orientation conventions (see \cref{tab:orientation}) --- if instead, \(n^a\) is taken to be future-directed on both \(\scri^\pm\) this reproduces the matching condition proposed by Strominger \cite{Stro-CK-match}.

From \cref{eq:charge-partial-matching}, we see that under the isomorphism \cref{eq:symm-match} we have
\be\label{eq:charge-matching}
    \mc Q[f^-; \nulls^-] \text{ along } \scri^- = \mc Q[f^+; \nulls^+] \text{ along } \scri^+
\ee
that is, the BMS-supermomenta on \(\scri^-\) and \(\scri^+\) are equal at \(i^0\) which is a direct consequence of the corresponding Spi-supertranslation \(\dd f\) on \(\cyl\) being totally fluxless. 

\begin{remark}[Change of conformal factor and rescaling function]
\label{rem:change-Sigma}
Our analysis is conformally-invariant on \(\scri\), while at \(i^0\), \(\omega\vert_{i^0} = 1\) and hence our analysis is independent of the choice of conformal factor. Thus it suffices to consider the change of the rescaling function \(\Sigma \mapsto \sigma \Sigma\) (\cref{rem:freedom-Sigma}) where \(\Lie_n \sigma = 0\) on \(\scri\). Under this change of rescaling function we have on \(\scri\)
\be
    n^a \mapsto n^a \eqsp l^a \mapsto l^a + \tfrac{1}{2} q^{ab} \nabla_b \ln\sigma + \tfrac{1}{8} q^{bc} \nabla_b \ln\sigma \nabla_c \ln\sigma ~ n^a
\ee
where the transformation of \(l^a\) follows from \cref{eq:L-defn,eq:l-defn}. Using \cref{eq:weyl-defn} we have
\begin{subequations}\begin{align}
    \begin{split}
    \Sigma^{-3} \mc P & \mapsto \sigma^{-3}\Sigma^{-3} \big[ \mc P + q^{ab} \mc S_a \nabla_b \ln\sigma + \tfrac{1}{4} q^{ac} q^{bd} \mc R_{ab} \nabla_c \ln\sigma \nabla_d \ln\sigma \big] \\
    &~ = \sigma^{-3} \big[ \Sigma^{-3} \mc P + \Sigma^{-1} \tilde q^{ab} \mc S_a \nabla_b \ln\sigma + \tfrac{1}{4}  \Sigma \tilde q^{ac} \tilde q^{bd} \mc R_{ab} \nabla_c \ln\sigma \nabla_d \ln\sigma \big] 
    \end{split} \\[1.5ex]
    \begin{split}
    \Sigma^{-1} \mc S_a & \mapsto \sigma^{-1} \Sigma^{-1} \lb[ \mc S_a + \tfrac{1}{2} q^{bc} \mc R_{ab} \nabla_c \ln\sigma \rb] \\
    &~ = \sigma^{-1} \lb[ \Sigma^{-1} \mc S_a + \tfrac{1}{2} \Sigma \tilde q^{bc} \mc R_{ab} \nabla_c \ln\sigma \rb]
    \end{split} \\[1.5ex]
    \Sigma \mc R_{ab} &\mapsto \sigma \Sigma \mc R_{ab}
\end{align}\end{subequations}
Since \(\sigma\) is \(C^{>-1}\) in null directions it follows from the above transformations that the falloff conditions \cref{eq:N-R-falloff,eq:S-falloff,eq:P-falloff} are preserved under a change of the rescaling function, and the field induced on \(\nulls^\pm\) by \(\lim_{\to i^0} \Sigma^{-3} \mc P\) transforms with rescaling weight \(r=-3\) as
\be
    (\Sigma^{-3} \mc P) \vert_{\nulls^\pm} \mapsto (\dd\sigma^\pm)^{-3} (\Sigma^{-3} \mc P) \vert_{\nulls^\pm}
\ee
where \(\dd\sigma^\pm\) is the function induced on \(\nulls^\pm\) by \(\lim_{\to i^0} \sigma\) along \(\scri^\pm\).
Similarly on \(\cyl\), using \cref{eq:U-defn}, we have
\be\begin{split}
    (\dd \Sigma^{-1} \dd E_{ab} \dd U^a \dd U^b)\vert_{\nulls^\pm} \mapsto (\dd\sigma^\pm)^{-3} (\dd \Sigma^{-1} \dd E_{ab} \dd U^a \dd U^b)\vert_{\nulls^\pm}
\end{split}\ee
which also has rescaling weight \(r = -3\), where we have used the fact that \(\sigma\) is \(C^{>-1}\) in both null and spatial directions so that on \(\cyl\), \(\lim_{\to \nulls^\pm} \dd \sigma = \dd\sigma^\pm\). Further, the area element \(\tilde{\dd \varepsilon}_{ab}\) induced on \(\nulls^\pm\) from both \(\scri\) and \(\cyl\) has rescaling weight \(r=2\), and the function \(\dd F^\pm\) representing the supertranslations has rescaling weight \(r=1\)
\be
    \tilde{\dd \varepsilon}_{ab} \mapsto (\dd \sigma^\pm)^2 \tilde{\dd \varepsilon}_{ab} \eqsp \dd F^\pm \mapsto \dd\sigma^\pm \dd F^\pm
\ee
Thus, the supermomenta on \(\nulls^\pm\) induced from both \(\scri\) and \(\cyl\), and our matching result are independent of the choice of rescaling function and conformal factor.
\end{remark}

\begin{remark}[Lorentz invariance in \(Ti^0\)]
\label{rem:lor-inv}
    Consider the action of a Lorentz transformation \(\Lambda \in SO(1,3)\) at \(i^0\). Under the action of this transformation on \(Ti^0\), a given unit-spatial direction \(\vec\eta \in \hyp\) is mapped to another unit-spatial direction \(\Lambda \circ \vec\eta \in \hyp\). Thus, Lorentz transformations act as isometries on the space of spatial directions \(\hyp\).\footnote{For any infinitesimal Lorentz transformation represented by a \emph{direction-independent} antisymmetric tensor \(\dd\Lambda_{ab}\) at \(i^0\), the corresponding \emph{direction-dependent} Killing field on \(\hyp\) is given by \(\dd X^a(\vec\eta) = \dd\varepsilon^{abc}(\vec\eta) \dd\Lambda_{bc}\) \cite{AH,Ash-in-Held}. } Note, that the choice of rescaling function used to construct \(\cyl\) need not be Lorentz-invariant, i.e., \(\Lambda \circ \Sigma \neq \Sigma\). However, any null or (rescaled) spatial direction \(\vec N\) is mapped to another direction \((\Lambda \circ \Sigma) (\Lambda \circ \vec N)\) which is also null or spatial respectively. Thus, the space \(\cyl\) constructed with some choice of rescaling function \(\Sigma\) is mapped to another space \(\cyl'\) constructed using the rescaling function \(\Sigma' = (\Lambda \circ \Sigma)\). Since we consider \(\cyl\) as an abstract manifold with a conformal-class (under a change of rescaling function) of metrics \(\tilde{\dd h}_{ab}\) as described in \cref{sec:cyl}, any Lorentz transformation \(\Lambda\) acts as a conformal isometry of \(\cyl\) with the change in rescaling function given by \(\sigma = \Sigma^{-1} (\Lambda \circ \Sigma)\). Further the action of any Lorentz transformation commutes with the reflection \(\Upsilon\). As we have already shown that our result is invariant under a change of the rescaling function, it follows that our matching result is also invariant under Lorentz transformations. 
\end{remark}

\begin{remark}[Invariance under logarithmic translations]
\label{rem:log-trans-inv}
    Consider the logarithmic translation ambiguity in the \(C^{>1}\)-structure at \(i^0\). As described in \cref{rem:log-trans} let \(x^i\) and \(\tilde x^i\) be two \emph{inequivalent} \(C^{>1}\)-charts at \(i^0\), related by the coordinate transformation \cref{eq:log-trans} and the unphysical metrics by \cref{eq:log-omega-factor}. Since both \(\partial x^i/\partial \tilde x^j\) and the relative conformal factor \(\tilde \omega\) are \(C^0\) at \(i^0\), the space of spatial directions \(\hyp\) is unaffected by logarithmic translations. Similarly, it can be verified that any choice of rescaling function \(\Sigma\) satisfies the properties \cref{def:Sigma} in the \(\tilde x^i\)-chart if it does so in the \(x^i\)-chart. Thus, the space of null and spatial directions \(\cyl\) is also unaffected by the logarithmic translations. Further, since \(\dd E_{ab}\) is invariant under the logarithmic translations \cite{Ash-log}, it follows that the matching of supertranslations and supermomenta is independent of the logarithmic translation ambiguity in the Ashtekar-Hansen structure. Note that since for null-regular spacetimes \(\dd E_{ab}\) is even under the reflection map \(\Upsilon\) (see \cref{sec:wave-hyp}), we can follow the analysis of \cite{Ash-log} to eliminate the logarithmic translation ambiguity by demanding that the potential \(\dd E\) (\cref{eq:E-potential}) is also reflection-even (for Minkowski spacetime this uniquely picks out the choice \(\dd E = 0\)). However this additional parity condition on the potential \(\dd E\) is not required for our result.
\end{remark}

\hr

With the diagonal symmetry algebra \(\mf s^\times\) we can now analyse the conservation of flux of BMS-supermomenta between \(\scri^-\) and \(\scri^+\). In any choice of the rescaling function \(\Sigma\), consider any totally fluxless supertranslation \([\dd F] \in \mf s^\times\), and let \(f^\pm \in \mf s^\pm\) be the unique BMS-supertranslations on \(\scri^\pm\) determined by the boundary values on \(\nulls^\pm\) of any representative \(\dd F \in [\dd F]\). Let \(S^\pm\) be some (finite) cross-sections of \(\scri^\pm\), respectively, and let \(\mc F^\pm\) denote the BMS-supermomentum flux between \(i^0\) and \(S^\pm\) corresponding to \(f^\pm\). Note that in our convention both \(\mc F^\pm\) are \emph{incoming} fluxes into the physical spacetime.

From the preceding analysis we know that for any totally fluxless supertranslation \([\dd F] \in \mf s^\times\) the corresponding BMS-supermomenta for \(f^\pm\) at \(i^0\) from both \(\scri^\pm\) match (\cref{eq:charge-matching}), and the fluxes \(\mc F^\pm\) are finite (\cref{rem:finite-flux-scri}). This immediately gives us the conservation law for the BMS-supermomenta defined on the cross-sections \(S^\pm\) of \(\scri^\pm\)
\be
    \mc Q[f^+; S^+] - \mc Q[f^-; S^-] = \mc F^+[f^+] + \mc F^-[f^-]
\ee 
Further, if \(\scri^\pm\) each have future/past boundaries at timelike infinities \(i^\pm\), respectively, and the spacetime satisfies appropriate conditions at \(i^\pm\) (see \cref{rem:timelike-infinity}) so that the BMS-supermomenta vanish as \(S^\pm \to i^\pm\), then we have the \emph{global} conservation law
\be\label{eq:global-conservation}
    \mc F^+[f^+; \scri^+] + \mc F^-[f^-; \scri^-] = 0
\ee
This implies that the total \emph{incoming} flux on \(\scri^-\) equals the total \emph{outgoing} flux at \(\scri^+\) and thus the flux is conserved in the scattering from \(\scri^-\) to \(\scri^+\).

\begin{remark}[Timelike infinities \(i^\pm\)]
\label{rem:timelike-infinity}
To obtain the global conservation law \cref{eq:global-conservation}, one needs suitable falloff conditions on the gravitational fields at the timelike infinities \(i^\pm\). However, unlike at \(i^0\) the behaviour of the fields at \(i^\pm\) is completely determined by the Einstein equations along with suitable initial data. Any falloff conditions one imposes at \(i^\pm\) should allow for, at the very least, solutions with initial data which is compactly-supported (in some suitable gauge), or more generally, initial data which is asymptotically-flat at \(i^0\) in the sense of \cref{def:AH} on a Cauchy surface. If one assumes that the spacetime becomes asymptotically-flat at \(i^\pm\), analogous to the conditions at \(i^0\) \cite{Porrill, Cutler}, one can adapt our method above to derive global conservation laws as in \cref{eq:global-conservation}. In these cases, for each \(i^\pm\), the analogues of the blowups \(\hyp\) and \(\cyl\) are a spacelike-unit-hyperboloid and a \(3\)-dimensional ball with a single boundary, respectively. However, in general relativity, even ``low frequency'' data on \(\scri^-\), in general, can lead to black hole formation \cite{Ch-BH}. In such spacetimes, the presence of an event horizon implies that the metric \(g_{ab}\) can only be assumed to be \(C^{>-1}\) at \(i^\pm\), and a more careful analysis is needed at the timelike infinities \cite{HL-timelike}. Further, one would also have to take into account the fluxes associated to symmetries on the event horizon \cite{CFP}.
\end{remark}

\section{Discussion}
\label{sec:disc}

We showed that the matching conditions for asymptotic supertranslations and supermomentum charges conjectured in \cite{Stro-CK-match} are satisfied by suitably regular asymptotically-flat spacetimes. The essential ingredients in our analysis are
\begin{enumerate*}
    \item the space \(\cyl\) of both null and spatial directions which allows us to simultaneously consider the limits of (suitably rescaled) Weyl tensor fields in both null and spatial directions at \(i^0\). 
    \item the reflection conformal isometry \(\Upsilon\) of \(\cyl\) (\cref{eq:reflection-cyl}) which provides an identification of the null generators of \(\scri^-\) to those on \(\scri^+\).
    \item the null-regularity conditions on the gravitational fields along null directions in \cref{def:null-regular} which ensure that the BMS-supermomenta defined on \(\scri\) admit limits as one approaches \(i^0\).
    \item the choice of the totally fluxless subalgebra of Spi-supertranslations on \(\cyl\), ensuring the physical criteria that there is no flux across spatial infinity in a scattering process.
\end{enumerate*}
The Einstein equations on \(\cyl\) near spatial infinity, then imply that the relevant parts of the asymptotic Weyl tensor from \(\scri^-\) to \(\scri^+\) antipodally match at \(i^0\) and reduce the supertranslation algebra to the diagonal subalgebra \(\mf s^\times\). As a consequence we showed that the BMS-supermomenta associated to \(\mf s^\times\) on \(\scri^\pm\) match at \(i^0\) and the corresponding fluxes are conserved.\\

The smoothness conditions in \cref{def:AH} can be weakened to allow for certain logarithmic terms in the asymptotic behaviour of the unphysical metric: at \(\scri\), \(g_{ab}\) can be polyhomogenous in the sense of \cite{CMS-poly} while at \(i^0\), \(g_{ab}\) can be \(C^{0^+}\) in the sense of \cite{Herb-dd}. These weaker conditions still allow for a sufficiently regular Weyl tensor so that the fields \(\mc R_{ab}, \mc S_a, \mc P\) at \(\scri\) (\cref{eq:weyl-defn}) and \(\dd E_{ab}\) on \(\hyp\) (\cref{eq:E-defn}) are well-defined. The rest of our analysis can then be carried out in the same manner. The null-regularity conditions (\cref{def:null-regular}) imposed on the gravitational fields are crucial in our analysis --- they ensure that the flux of charges radiated through \(\scri^\pm\) in the scattering process are finite (\cref{rem:finite-flux-scri}). At spatial infinity these conditions discard the solutions for which the electric part of the asymptotic Weyl tensor \(\dd E_{ab}\) is odd under the reflection \(\Upsilon\) (see \cref{sec:wave-hyp}). The supermomenta for such solutions on the space of null directions \(\nulls^\pm\) diverges and the flux of supermomenta radiated through null infinity will not be finite; thus such solutions are unphysical from the point of view of a scattering problem. It can be verified that our null-regularity conditions are automatically satisfied when the unphysical metric \(g_{ab}\) is \(C^{>0}\) in both null and spatial directions at \(i^0\) --- the Kerr family of spacetimes falls into this class as shown by Herberthson \cite{Herb-Kerr}, and also under the weaker assumption that \(X^a \nabla_a\) is \(C^{>-1}\) for any vector field \(X^a\) which is tangent to \(\scri\) and \(C^{>-1}\) at \(i^0\) along null directions, that is, one only requires that the derivatives tangent to \(\scri\) admit regular direction-dependent limits to \(i^0\) \cite{HL-GR-matching}. They also hold when the unphysical metric is only \(C^{0^+}\) at \(i^0\) in the sense of \cite{Herb-dd}.

While we have argued that the conditions in \cref{def:AH,def:null-regular} are physically reasonable, an important question we have left unanswered is the existence of a ``large enough'' class of solutions of the Einstein equation which satisfy these conditions. To resolve this, one would have to show that there exist initial data --- with some appropriate topology and suitable falloffs near \(i^0\) ---  either on a spatial Cauchy surface or on \(\scri^-\), which evolve through the Einstein equation to a spacetime satisfying \cref{def:AH,def:null-regular}. It seems this issue is best analysed in the formalism of Friedrich \cite{Friedrich} which also involves a cylindrical blowup of spatial infinity, similar to the space \(\cyl\) used in our analysis. The analysis of \cite{Tro} suggests that there do exist many solutions of the linearised Einstein equation on Minkowski spacetime satisfying \cref{def:null-regular}. However, in the conformal-completion used by Friedrich the unphysical metric does not have a limit to \(i^0\) along spatial directions, even for Minkowski spacetime (see \cite{Tro}), and the relationship of this formalism to that of Ashtekar and Hansen is not clear, but certainly merits further investigation.

We can also compare the falloffs implied by \cref{def:null-regular} to those in the class of spacetimes constructed in the global nonlinear stability analysis of Minkowski spacetime; we use the Bondi-Sachs conformal frame and the Bondi-Sachs parameter \(u\) described in \cref{sec:BS}. In the Christodoulou-Klainerman (CK) class of spacetimes \cite{CK}, one has \(N_{ab}^\bs = O(1/u^{\nfrac{3}{2}})\) and \(\mc P^\bs\) becomes spherically-symmetric near \(i^0\) as fast as \(1/u^{\nfrac{1}{2}}\). Thus, CK-spacetimes have finite Bondi mass at \(i^0\) but all other the BMS-supermomenta vanish \cite{Ash-CK-triv}. The conditions used by Bieri \cite{Bieri-CK-ext-thesis,Bieri-CK-ext0,Bieri-CK-ext1} allow \(N_{ab}^\bs = O(1/u^{\nfrac{1}{2}})\), while the slightly stronger conditions used by Bieri and Chru\'sciel \cite{BC-mass} give \(N_{ab}^\bs = O(1/u^{\nfrac{1}{2}+\epsilon})\) (which is also the falloff used in \cite{Ash-Mag-Ash}). In the latter case, the Bondi mass has a limit to \(i^0\) and coincides with the ADM mass, but due to the slow falloff of the News tensor we expect that general BMS-supermomenta do not admit a limit to \(i^0\) in such spacetimes. Our assumed falloff on the News tensor \(N_{ab}^\bs = O(1/u^{1+\epsilon})\) (\cref{eq:null-regular-BS}) lies between the cases of Christodoulou-Klainerman and Bieri-Chru\'sciel, and allow for a finite yet non-vanishing BMS-supermomenta at \(i^0\). We expect our conditions to be compatible with another intermediate case considered by Bieri \cite{Bieri-pp} (denoted by (B2) in \cite{Bieri-magnetic}).\\

The analogous analysis for the charges associated to the full BMS group is trickier. Neither the BMS group nor the Spi group contain a unique Lorentz subgroup or a unique Poincar\'e subgroup. This results in the well-known problem of supertranslation ambiguities in defining the charges for a Lorentz subalgebra, such as angular momentum, on null infinity and at spatial infinity. For stationary spacetimes the angular momentum can be defined at the cost of reducing the BMS algebra to the Poincar\'e algebra \cite{NP-red}. Similarly at spatial infinity, the angular momentum can be defined by imposing, either the Regge-Teitelboim  parity conditions \cite{RT-parity} in a \((3+1)\)-formalism, or, stronger falloff conditions on the magnetic part of the Weyl tensor at \(i^0\) in the Ashtekar-Hansen framework \cite{AH}, again reducing the asymptotic symmetry algebra at \(i^0\) to the Poincar\'e algebra. In this case, the analysis of \cite{AS-ang-mom} proves the matching conditions for angular momentum (and other Lorentz charges). We defer the detailed investigation of this problem in the general case to future work.

\section*{Acknowledgements}
I would like to thank \'Eanna \'E. Flanagan for suggesting this problem and helpful discussions throughout this work. I would also like to thank Ho DM for some entertaining help with editing. This work is supported in part by the NSF grants PHY-1404105 and PHY-1707800 to Cornell University.

\appendix

\section{Null infinity in arbitrary conformal choices}
\label{sec:non-Bondi}

In this section we obtain the evolution equations of the Weyl tensor, and expressions for the News tensor and supertranslations on \(\scri\) in an arbitrary choice of conformal factor \(\Omega\) which is not restricted to satisfy the Bondi condition i.e. \(\nabla_a \nabla_b \Omega \neq 0\) on \(\scri\). The evolution of the Weyl tensor on \(\scri\) can be directly obtained using the GHP-formalism. For the rest, it will be convenient to proceed in the following manner: We first write the relevant expressions in the GHP-formalism in the Bondi-Sachs conformal frame i.e., where the conformal factor \(\Omega_\bs\) is chosen such that
\be
    \nabla_a^\bs \nabla_b^\bs \Omega_\bs \vert_\scri = 0 \eqsp q_{ab}^\bs = s_{ab}
\ee
where \(s_{ab}\) is the unit-metric on \(\bb S^2\). Then we use the conformally-covariant version of the GHP-derivative operators to generalise these expressions to an arbitrary conformal choice. Finally, we convert these expressions from the GHP notation to the tensor notation used in the main body of the paper.

We begin with summarising the GHP-formalism (for details see \cite{NP,GHP}). To conform to our conventions for the signature and Reimann curvature of the metric we will use the sign conventions of \cite{SK}.\footnote{In the notation of \cite{SK}, their \(l^a\) corresponds to our \(n^a\) and their \(k^a\) corresponds to our \(l^a\).} We pick a null tetrad \((n^a, l^a, m^a, \bar m^a)\) normalised so that \( n_a l^a = - m_a \bar m^a = - 1 \) and all other inner products vanish. The metric can then be written as \(g_{ab} = - 2 n_{(a} l_{b)} + 2 m_{(a} \bar m_{b)}\). Given a metric the freedom in the choice of null tetrad is given by the \emph{GHP-transformations}
\be\label{eq:GHP-transform}
	l^a \mapsto \lambda \bar\lambda l^a \eqsp n^a \mapsto (\lambda \bar\lambda)^{-1} n^a \eqsp m^a \mapsto \lambda (\bar\lambda)^{-1} m^a
\ee
for any complex scalar \(\lambda\). A scalar field \(\xi\), associated to the choice of null tetrad, has a \emph{GHP-weight} of \((p,q)\) if under the GHP-transformations \cref{eq:GHP-transform} \(\xi\) transforms as
\be\label{eq:GHP-wt-defn}
    \xi \mapsto \lambda^p \bar\lambda^q \xi
\ee
We will denote the GHP-weight of such fields as \(\xi \wt (p,q)\).

We define the GHP spin coefficients by (following the conventions of \cite{SK}, these differ by a sign from \cite{NP,GHP})
\begin{subequations}\label{eq:spin-defn}\begin{align}
    - \kappa & \defn m^a l^b \nabla_b l_a \wt (3,1) \eqsp & - \rho & \defn m^a \bar m^b \nabla_b l_a \wt (1,1) \\
    - \sigma & \defn m^a m^b \nabla_b l_a \wt (3,-1) \eqsp & - \tau & \defn m^a n^b \nabla_b l_a \wt (1,-1) \\
    - \kappa' & \defn \bar m^a n^b \nabla_b n_a \wt (-3,-1) \eqsp & - \rho' & \defn \bar m^a m^b \nabla_b n_a \wt (-1,-1) \\
    - \sigma' & \defn \bar m^a \bar m^b \nabla_b n_a \wt (-3,1) \eqsp & -\tau' & \defn \bar m^a l^b \nabla_b n_a \wt (-1,1) \\[1.5ex]
    - \epsilon & \defn \tfrac{1}{2} (n^a l^b \nabla_b l_a - \bar m^a l^b \nabla_b m_a) \eqsp & -\beta & \defn \tfrac{1}{2} (n^a m^b \nabla_b l_a - \bar m^a m^b \nabla_b m_a) \\
    - \epsilon' & \defn \tfrac{1}{2} (l^a n^b \nabla_b n_a - m^a n^b \nabla_b \bar m_a) \eqsp & -\beta' & \defn \tfrac{1}{2} (l^a \bar m^b \nabla_b n_a - m^a \bar m^b \nabla_b \bar m_a) 
\end{align}\end{subequations}
The spin coefficients \(\epsilon, \beta, \epsilon', \beta'\) are not GHP-weighted quantities (their transformations contain derivatives of \(\lambda\) appearing in \cref{eq:GHP-transform}), but instead define the \emph{GHP derivative operators} \((\thorn, \thorn',\eth,\eth')\) as follows 
\be\label{eq:GHP-d}\begin{aligned}
	\thorn\xi & \defn (l^a \nabla_a - p\epsilon - q \bar\epsilon)\xi &&\wt (p+1,q+1) \\
	\thorn'\xi & \defn (n^a \nabla_a + p\epsilon' + q \bar\epsilon')\xi &&\wt (p-1,q-1) \\
	\eth\xi & \defn (m^a \nabla_a - p\beta + q \bar\beta')\xi &&\wt (p+1,q-1) \\
	\eth'\xi & \defn (\bar m^a \nabla_a + p\beta' - q \bar\beta)\xi &&\wt (p-1,q+1)
\end{aligned}\ee
where \(\xi \wt (p,q)\) is any GHP-weighted scalar. 

We also define the Weyl scalars in the chosen tetrad by
\be\label{eq:Weyl-GHP}\begin{aligned}
    \Psi_4 & \defn C_{abcd} \bar m^a n^b \bar m^c n^d &&\wt (-4,0) \\
    \Psi_3 & \defn C_{abcd} l^a n^b \bar m^c n^d &&\wt (-2,0) \\
    \Psi_2 & \defn \tfrac{1}{2} C_{abcd} (l^a n^b l^c n^d - l^a n^b m^c \bar m^d) && \\
    & ~= C_{abcd} l^a m^b \bar m^c n^d &&\wt (0,0) \\
    \Psi_1 & \defn C_{abcd} l^a m^b l^c n^d &&\wt (2,0) \\
    \Psi_0 & \defn C_{abcd} l^a m^b l^c m^d &&\wt (4,0)
\end{aligned}\ee
The relations between the spin coefficients, Weyl and Ricci tensors in the GHP-formalism can be found in \cite{GHP}.

\hr

Now we adapt the choice of the null tetrad in the unphysical spacetime to \(\scri\). Since the conformal factor \(\Omega\) is independent of the choice of a null tetrad we have the GHP-weights \(\nabla^a\Omega \wt (0,0)\). To use the GHP formalism consistently with the transformations \cref{eq:GHP-transform} we define on \(\scri\)
\be\label{eq:A-defn}
    An^a = \nabla^a\Omega \quad\text{where } A \wt(1,1)
\ee
and in the end we shall set \(A=1\). As in \cref{sec:AH}, we choose \(l^a\) to be the auxilliarly null normal to some choice of foliation of \(\scri\). Then the remaining tetrads \((m^a, \bar m^a)\) are a \emph{complex} orthonormal basis for the cross-sections of \(\scri\) of the chosen foliation, and we have
\be
    q_{ab} = 2 m_{(a} \bar m_{b)} \eqsp \varepsilon_{ab} = 2 i m_{[a} \bar m_{b]}
\ee

Using \(\nabla_a \nabla_b \Omega\vert_\scri = \Phi g_{ab} \) (from \cref{eq:n-Phi}) and \cref{eq:A-defn} in \cref{eq:spin-defn} we get on \(\scri\)\footnote{Note that \cref{eq:spin-scri} also implies that \((\thorn'+\rho')A = \eth A = 0\) which is Eq.~9.8.26 \cite{PR2}.}
\be\label{eq:spin-scri}\begin{aligned}
    \kappa' &= \sigma' = \tau' = 0 \eqsp & \rho' & = - A^{-1}\Phi \\[1.5ex]
    \epsilon &= - \tfrac{1}{2} (n^a l^b \nabla_b l_a - \bar m^a l^b \nabla_b m_a) \eqsp & \beta & = \tfrac{1}{2}( m^a \nabla_a \ln A + \bar m^a m^b \nabla_b m_a ) \\
    \epsilon' &= \tfrac{1}{2} (- \Lie_n \ln A + A^{-1}\Phi + m^a n^b \nabla_b \bar m_a) \eqsp & \beta' & = \tfrac{1}{2}( - \bar m^a \nabla_a \ln A + m^a \bar m^b \nabla_b \bar m_a )
\end{aligned}\ee
Further since \(l_a\) is normal to the cross-sections of \(\scri\) we have \(\rho = \bar \rho\) (i.e., the twist of \(l_a\) vanishes), while \(\kappa, \sigma, \tau\) are arbitrary. The spin coefficients \(\tau\) and \(\sigma\) are related to the tensors \(\tau_a\) and \(\sigma_{ab}\) defined in \cref{eq:tau-defn,eq:sigma-defn} by
\be\label{eq:tau-sigma-GHP}\begin{aligned}
    \tau_a & = - \tau \bar m_a - \bar \tau m_a  \eqsp& \tau & = - \tau_a m^a \\
     \sigma_{ab} & = - \sigma \bar m_a \bar m_b - \bar\sigma m_a m_b \eqsp& \sigma & = - \sigma_{ab} m^a m^b \\
\end{aligned}\ee

By the peeling theorem all the Weyl scalars \(\Psi_i\) (\(i = 0, \ldots, 4\)) defined in \cref{eq:Weyl-GHP} vanish at \(\scri\), and \(\psi_i \defn \Omega^{-1}\Psi_i\) has a limit to \(\scri\) (see Theorem~9.6.41 \cite{PR2}). These are related to the tensor fields defined in (\cref{eq:weyl-defn}) by
\be\label{eq:weyl-GHP}
    \psi_4 = \mc R_{ab}\bar m^a \bar m^b \eqsp \psi_3 = \mc S_a \bar m^a \eqsp \psi_2 = \tfrac{1}{2} (\mc P + i \mc P^*) \eqsp \psi_1 = \mc J_a m^a \eqsp \psi_0 = \mc I_{ab} m^a m^b \\
\ee
Note that each \(\psi_i\) has the same GHP-weight as the corresponding \(\Psi_i\) given in \cref{eq:Weyl-GHP}.

We first note some relations on \(\scri\) which are independent of the choice of the conformal factor \(\Omega\) and can be directly obtained from the GHP equations in \cite{GHP}. \cref{eq:curl-weyl} implies that the Weyl scalars \(\psi_i\) on the unphysical spacetime satisfy the same Bianchi identity at \(\scri\) as the Weyl scalars \(\hat\Psi_i\) defined with the physical Weyl tensor with all the physical Ricci tensor terms set to vanish by the vacuum Einstein equations (see Eqs.~9.10.1 and 9.10.2 \cite{PR2}). Taking the GHP-prime of the Bianchi identities Eqs.~2.33--2.36 \cite{GHP}, with their \(\Psi_i\) replaced by \(\psi_i\) and the Ricci tensor terms set to vanish, we have
\begin{subequations}\label{eq:weyl-evol-GHP}\begin{align}
    (\thorn' - 4\rho') \psi_3 & = (\eth - \tau) \psi_4 \\
    (\thorn' - 3\rho') \psi_2 & = (\eth - 2\tau) \psi_3 + \sigma \psi_4 \\
    (\thorn' - 2\rho') \psi_1 & = (\eth - 3 \tau) \psi_2 + 2 \sigma \psi_3 \\ 
    (\thorn' - \rho') \psi_0 & = (\eth - 4 \tau) \psi_1 + 3 \sigma \psi_2 
\end{align}\end{subequations}
If the conformal factor satisfies the Bondi condition i.e. \(\rho' = 0\), these reduce to Eqs.~9.10.4--9.10.7 \cite{PR2}. Using \cref{eq:weyl-GHP}, along with \cref{eq:GHP-d,eq:spin-scri} and setting \(A=1\), \cref{eq:weyl-evol-GHP} is equivalent to \cref{eq:weyl-evol}. Further, from Eq.~2.25 \cite{GHP} and \cref{eq:spin-scri} we have
\be\label{eq:News-Ric-GHP}
    - \thorn' \sigma = \Phi_{02} - \rho' \sigma - (\eth - \tau) \tau 
\ee
where, in our conventions, \(\Phi_{02} \defn \tfrac{1}{2} S_{ab} m^a m^b\) \cite{SK}. \\

The expressions which lead to \cref{eq:weyl-News} can also be derived directly from the GHP-formalism following the computations described in \S~9.8 \cite{PR2} but keeping a general conformal factor. However, it will be easier to use the already available form of these expressions in a Bondi-Sachs conformal frame and then generalise them to arbitary conformal frames. In a Bondi-Sachs frame the \emph{complex News function} \(N\) is defined by (Eq.~9.8.75 \cite{PR2})
\be\label{eq:News-defn-BS}
    \bar N \defn - \thorn' \sigma \qquad\text{(Bondi-Sachs)}
\ee
which is related to the Weyl scalars through (Eqs.~9.8.82 and 9.8.83 \cite{PR2})
\be\label{eq:weyl-News-BS}
    A \psi_4 = \thorn' N \eqsp A \psi_3 = \eth N \qquad\text{(Bondi-Sachs)}
\ee

Our goal is to generalise \cref{eq:News-defn-BS,eq:weyl-News-BS} to arbitrary conformal choices. Under a change of conformal factor \(\Omega \mapsto \omega \Omega\), recall the transformations of the metric and null tetrad on \(\scri\) 
\be\label{eq:conf-g-tetrad}
    g_{ab} \mapsto \omega^2 g_{ab} \eqsp (n^a, l^a, m^a, \bar m^a)  \mapsto \omega^{-1} (n^a, l^a, m^a, \bar m^a)
\ee
with the GHP-weights \(\omega \wt (0,0)\). Under this transformation a scalar field \(\xi\) associated to the choice of null frame we have the additional conformal weight defined in \cref{eq:conf-wt-defn}. We will denote the combined \emph{conformal-GHP-weight} of such scalars as \(\xi \wt (p,q; w)\).

Under a conformal transformation \cref{eq:conf-g-tetrad}, the GHP-derivatives of a conformally-weighted GHP-scalar \(\xi\) (\cref{eq:GHP-d}) will, in general, pick up derivatives of \(\omega\). However, the spin coefficients \(\rho'\) and \(\tau\) are not conformally-weighted since (see also \cref{eq:conf-trans})
\be
    \rho' \mapsto \omega^{-1} (\rho' - \thorn' \ln\omega) \eqsp \tau \mapsto \omega^{-1} (\tau - \eth \ln\omega)
\ee
Thus, we can ``correct'' the conformal behaviour of the GHP-derivatives by adding suitable combinations of \(\rho'\) and \(\tau\) to define the \emph{conformal-GHP-derivatives}. For our purposes we only need the conformal-GHP-derivatives tangential to \(\scri\) which are given by (see Eq.~5.6.36 \cite{PR1}, where we use \(w_0 = w_1 = -\half\) as in see Eq.~5.6.26 (iii) \cite{PR1} to be compatible with the conformal transformations of the null tetrad \cref{eq:conf-g-tetrad} )%
\footnote{We note that the spin coefficient \(\tau'\) is, in fact, conformally-invariant on \(\scri\) contrary to its transformation given in Eq.~5.6.27 (iii) \cite{PR1}. This difference arises since under \(\Omega \mapsto \omega \Omega\) the normal \(n_a\) transforms away from \(\scri\) as \(n_a \mapsto \omega n_a + \Omega A \nabla_a \omega\). For this reason we have defined the conformal-GHP-derivative corresponding to \(\eth'\) with a \(\bar\tau\) instead of \(\tau'\). This corresponds to the conformal operators \(\tfrac{1}{2}(\eth_{\ms C}+\bar\eth'_{\ms C})\) and \(\tfrac{1}{2}(\bar\eth_{\ms C} + \eth'_{\ms C})\) introduced in Eq.~5.6.36 \cite{PR1}.}
\be\label{eq:conf-GHP-d}\begin{split}
    \lb[ \thorn' + (w + \tfrac{1}{2}(p+q))\rho' \rb]\xi &\wt (p-1,q-1; w-1) \\
    \lb[ \eth + (w - \tfrac{1}{2}(p-q))\tau\rb]\xi &\wt (p+1,q-1; w-1) \\
    \lb[ \eth' + (w + \tfrac{1}{2}(p-q))\bar\tau \rb]\xi &\wt (p-1,q+1; w-1)
\end{split}\ee
where \(\xi \wt (p,q; w) \) is any conformally-weighted GHP-scalar. In a Bondi-Sachs frame \(\rho' = \tau = 0\) and so the conformal-GHP-derivatives \cref{eq:conf-GHP-d} are equivalent to the usual GHP-derivatives. Thus, to generalise any expression in the Bondi-Sachs frame to arbitrary conformal choices, we can replace the usual GHP-derivatives in that expression with the conformal-GHP-derivatives taking into account the appropriate weights. \cref{eq:weyl-evol-GHP} is already in a conformal-invariant form where each \(\psi_i\) has conformal weight \(w = -3\), and the conformal-invariance of \cref{eq:News-Ric-GHP} can be verified though the computation is tedious.

With this setup we return to \cref{eq:News-defn-BS,eq:weyl-News-BS}. The spin coefficient \(\sigma\) is conformal-GHP-weighted according to \(\sigma \wt (3,-1; -1) \). Replacing the \(\thorn'\) in \cref{eq:News-defn-BS} by the corresponding conformal-GHP operator in \cref{eq:conf-GHP-d} the conformal and GHP-weights cancel. Thus \cref{eq:News-defn-BS} holds in any conformal frame, giving\footnote{Note that \cref{eq:News-sigma-conf} differs from the definition of the News function in Eq.~9.8.73 \cite{PR2} by terms involving the spin coefficient \(\tau\), which vanishes in the Bondi-Sachs frames. In arbitrary conformal frames we choose \cref{eq:News-sigma-conf} as the definition of the News function, since the corresponding News tensor \cref{eq:News-GHP} is conformally-invariant.}
\be\label{eq:News-sigma-conf}
    \bar N = - \thorn' \sigma \wt (2,-2; -2)
\ee
The \emph{conformally-invariant} News tensor \(N_{ab}\) is given by
\be\label{eq:News-GHP}
    N_{ab} = 2 N m_a m_b + 2 \bar N \bar m_a \bar m_b \eqsp 2 \bar N = N_{ab} m^a m^b 
\ee
Using \cref{eq:tau-sigma-GHP,eq:News-Ric-GHP,eq:News-sigma-conf,eq:News-GHP} we reproduce the definition of the News tensor in \cref{eq:News-defn} and the relation in \cref{eq:News-Ric}.

The conformal-GHP-weights of the News function are \(N \wt (-2,2; -2) \), and again replacing the \(\thorn'\) and \(\eth\) in \cref{eq:weyl-News-BS} by the corresponding operators \cref{eq:conf-GHP-d}, we get
\be\label{eq:weyl-News-conf}
   A \psi_4 = (\thorn' - 2\rho') N \eqsp A \psi_3 = \eth N
\ee
Rewriting the above in terms of tensors using \cref{eq:weyl-GHP,eq:News-GHP} and setting \(A = 1\) we get \cref{eq:weyl-News}.\\

We can similarly obtain expressions for the BMS-supertranslations on \(\scri\) in arbitrary conformal choices. Let the BMS-supertranslation vector field for \(f\in \mf s^\pm\) on \(\scri\) be \(\xi^a = f \nabla^a \Omega = (fA) n^a\), where \(fA \wt (1,1; 1)\). In the Bondi-Sachs frame, we have 
\be
    \begin{split}
    \thorn' (fA) = 0 \quad\text{for } f \in \mf s^\pm  \\
    \eth^2 (f A) = 0 \quad\text{for } f \in \mf t^\pm
    \end{split}
    \qquad\text{(Bondi-Sachs)}
\ee
Carrying out the replacements of the GHP-derivatives as before, we have in arbitrary conformal frames
\be\begin{split}
    (\thorn' + 2\rho') (fA) = 0 \quad\text{for } f \in \mf s^\pm  \\
    (\eth - \tau)(\eth + \tau) (fA) = 0 \quad\text{for } f \in \mf t^\pm
\end{split}\ee
Using \cref{eq:GHP-d,eq:spin-scri} and setting \(A=1\) we reproduce \cref{eq:st-null,eq:trans-null}.\\

\begin{remark}[Universal structure and supertranslations at \(\scri\)]
\label{rem:univ-strc}
Note that the above procedure does not give us the boundary condition \(f\vert_{i^0} = 0\) on the BMS-supertranslations. This boundary condition can be obtained by transforming between the Bondi conformal frames to a general conformal choice (see \cref{eq:st-BS-to-gen}), or directly from the universal structure on \(\scri\) induced by asymptotic-flatness as follows: The universal structure at \(\scri\) (that is, the common structure induced on \(\scri\) by \emph{all} physical spacetimes satisfying \cref{def:AH}) is the pair \((n^a, q_{ab})\) where two pairs \((n^a, q_{ab})\) and \((\omega^{-1}n^a, \omega^2 q_{ab})\) are equivalent for any \(\omega\) satisfying the conditions in \cref{rem:freedom-Omega}. The infinitesimal diffeomorphisms generated by vector fields \(\xi^a\) on \(\scri\) which preserve this universal structure satisfy 
\be\label{eq:symm-null}
    \Lie_\xi n^a = - \alpha_{(\xi)} n^a \eqsp \Lie_\xi q_{ab} = 2 \alpha_{(\xi)} q_{ab}
\ee
for some function \(\alpha_{(\xi)}\) which depends on \(\xi^a\), is smooth on \(\scri\), and \(\alpha_{(\xi)} \vert_{i^0} = 0\) (since \(\omega\vert_{i^0} = 1\)). For BMS-supertranslations we are interested in vector fields of the form \(\xi^a = f n^a\) on \(\scri\) and evaluating \cref{eq:symm-null} for such vector fields (using \cref{eq:Lie-n-q}) we get
\be\label{eq:alpha-st}
    \alpha_{(\xi)} = \Lie_n f = \Phi f
\ee
This immediately tells us that \((\Lie_n -\Phi)f = 0\) on \(\scri\) and \(f\vert_{i^0} = 0\) which is \cref{eq:st-null}.\\
\end{remark}

\section{\(C^{>1}\) differential structure, direction-dependent tensors and space of directions at \(i^0\)}
\label{sec:dd}

Consider a manifold \(M\) which is smooth everywhere except at a point \(p \in M\) where it is \(C^1\) --- so that the tangent space \(Tp\) at \(p\) is well-defined. A function \(f\) in \(M\) is \emph{direction-dependent} at \(p\) if the limit of \(f\) along any curve \(\Gamma\), which is \(C^1\) at \(p\), exists and depends only on the tangent direction to \(\Gamma\) at \(p\). We write this as \(\lim_{\to p} f = \dd f(\vec N)\) where \(\vec N\) is the direction of the tangent to \(\Gamma\) at \(p\). Note that we can consider the limit \(\dd f(\vec N)\) as a function in the tangent space \(Tp\) which is constant along the rays (represented by \(\vec N\)) from \(p\).

\begin{definition}[Regular direction-dependent function \cite{Herb-dd}]
Let \(x^i\) denote a \(C^1\) coordinate chart with \(x^i(p) = 0\). A direction-dependent function \(f\) is \emph{regular direction-dependent} at \(p\) (with respect to the chosen chart \(x^i\)) if for all \(k\)
\be\label{eq:dd-derivative}
    \lim_{\to p} \lb( x^{i_1} \frac{\partial}{\partial x^{i_1}} \rb) \cdots \lb( x^{i_k} \frac{\partial}{\partial x^{i_k}} \rb) f = \lb[ \lb( x^{i_1} \frac{\partial}{\partial x^{i_1}} \rb) \cdots \lb( x^{i_k} \frac{\partial}{\partial x^{i_k}} \rb) \dd f \rb] (\vec N)
\ee
where on the right-hand-side we consider \(\dd f\) as a function in \(Tp\) as mentioned above. 
\end{definition}
\cref{eq:dd-derivative} ensures that, in the limit, \(f\) is smooth in its dependence on the directions \(\vec N\) --- the additional factors of \(x^i\) arise from converting the derivatives with respect to the ``rectangular'' coordinates \(x^i\) to derivatives with respect to the different directions (in \cref{eq:dd-der-spatial}, these factors are provided instead by \(\Omh\) \cite{AH,Ash-in-Held,Geroch-asymp}).

In general, the notion of regular direction-dependent tensors in any two coordinate charts in the same \(C^1\)-structure  are not equivalent \cite{AH, Herb-dd}. Thus, we restrict the differential structure of \(M\) at \(p\) to a \(C^{>1}\)-structure defined as follows.

\begin{definition}[\(C^{>1}\) differential structure]
\label{def:C>1-struc}
Consider any two \(C^1\) coordinate charts \(x^i\) and \(y^i\), in the same \(C^1\)-structure, containing the point \(p \in M\) such that for all \(i,j,k\) the \emph{transition functions}
\be\label{eq:C>1-struc}
    \frac{\partial^2 y^i(x)}{\partial x^j \partial x^k} \eqsp \frac{\partial^2 x^i(y)}{\partial y^j \partial y^k}
\ee
are regular direction-dependent at \(p\) in their respective coordinate charts. A collection of all coordinate charts related by \cref{eq:C>1-struc} defines a choice of \(C^{>1}\)-structure on \(M\) at \(p\).
\end{definition}

Given such a \(C^{>1}\)-structure at \(p\), any function whose derivatives upto the \((k-1)\)\textsuperscript{th} order vanish, whose \(k\)\textsuperscript{th} derivative is direction-independent, and whose \((k+1)\)\textsuperscript{th} derivative is regular direction-dependent will be called \(C^{>k}\). By a slight abuse of notation we denote regular direction-dependent functions by \(C^{>-1}\). Similarly, any tensor field is \(C^{>k}\) at \(p\) if all of its components in any coordinate chart in the chosen \(C^{>1}\)-structure are \(C^{>k}\) functions at \(p\).

\hr

Now we consider the \(C^{>1}\)-structure at the point \(p = i^0\) representing spatial infinity as defined in \cref{def:AH} and summarise the structure on the spaces of directions \(\hyp\) and \(\cyl\) (for details see Appendix~B \cite{KP-EM-match}). Since the metric \(g_{ab}\) in the Ashtekar-Hansen structure is universal at \(i^0\), it induces a metric \(\dd g_{ab}\) in the tangent space \(Ti^0\) which is isometric to the Minkowski metric. Thus, we can introduce \(C^{>1}\)-coordinates \(x^i = (t,x,y,z)\) in \(Ti^0\) so that
\be\label{eq:g-i0}\begin{split}
    \dd g_{ab} & \equiv - dt^2 + dx^2 + dy^2 + dz^2 \\
    & = - dt^2 + dr^2 + r^2 s_{AB} d\theta^A d\theta^B
\end{split}\ee
where the ``polar'' coordinates \((t,r,\theta^A)\) with \(\theta^A = (\theta, \phi)\) are defined in the usual way from \((t,x,y,z)\), and \(s_{AB}\) is the unit sphere metric
\be
    s_{AB} d\theta^A d\theta^B \equiv d\theta^2 + \sin^2\theta d\phi^2
\ee 

Let \(\alpha \defn t/r\), so that the spatial directions in \(Ti^0\) correspond to \(-1 < \alpha < 1\) while \(\alpha = \pm 1\) corresponds to null directions. The conformal factor can be chosen to be \(\Omega = r^2 - t^2\) so that
\be\label{eq:n-Ti0}
    n^a \defn \nabla^a \Omega \equiv 2r \lb( \alpha\partial_t + \partial_r \rb) 
\ee
while the \(C^{>-1}\) vector field
\be\label{eq:eta-Ti0}
    \dd\eta^a \defn \nabla^a \Omh \equiv \lb( 1 - \alpha^2 \rb)^{-\half} \lb( \alpha \partial_t + \partial_r \rb) 
\ee
represents the unit spatial directions \(\vec\eta\) in \(Ti^0\). The space of spatial directions is the unit-hyperboloid \(\hyp\) in \(Ti^0\) given by the surface \(r^2 - t^2 = 1\) with the induced metric\footnote{The forms of the metric on \(\hyp\) in \cref{eq:h-hyp-tau,eq:h-hyp-coord} are related by the coordinate change \(\alpha = \tanh \tau\).}
\be\label{eq:h-hyp-coord}
    \dd h_{ab} \equiv - \frac{1}{(1-\alpha^2)^2} d\alpha^2 + \frac{1}{1-\alpha^2} s_{AB} d\theta^A d\theta^B
\ee
A convenient choice of the rescaling function is given by.
\be\label{eq:Sigma-r}
    \Sigma^{-1} = r
\ee
On \(\hyp\) the function \(\dd \Sigma\) induced by \(\lim_{\to i^0}\Omh\Sigma\) in spatial directions is
\be\label{eq:Sigma-hyp}
    \dd \Sigma = \lb( 1 - \alpha^2 \rb)^\half 
\ee

The rescaled null and spatial directions \(\vec N\) are given by
\be\label{eq:N-Ti0}\begin{aligned}
    N^a = \tfrac{1}{2} \Sigma n^a & \equiv \lb( \alpha \partial_t + \partial_r \rb) \\
    & \equiv \lb( \pm \partial_t + \partial_r \rb) &&\quad \text{in null directions} \\
    & = \dd\Sigma \dd\eta^a &&\quad \text{in spatial directions}
\end{aligned}\ee

The cylinder \(\cyl\) is then given by the surface \(\Sigma^{-1 } = r = 1\) which is diffeomorphic to a conformal-completion of \(\hyp\) (with conformal factor \(\dd \Sigma\)), and the boundaries \(\nulls^\pm \cong \bb S^2\) correspond to \(\alpha = \pm 1\) representing the space of null directions at \(i^0\). With this choice the vector field \(\dd U^a\) and metric \(\tilde{\dd q}_{ab}\) on \(\nulls^\pm\) are
\be\label{eq:U-q-coord}
    \dd U^a \equiv \alpha (1-\alpha^2)^\half \partial_\alpha  \eqsp \tilde{\dd q}_{ab} \equiv s_{AB} d\theta^A d\theta^B
\ee

The reflection conformal isometry of \(\cyl\), \cref{eq:reflection-cyl}, is the map
\be
    \Upsilon : (\alpha, \theta^A) \mapsto (-\alpha, - \theta^A)
\ee
with \(\dd\varsigma = 1\) since our choice of \(\dd\Sigma\) (\cref{eq:Sigma-hyp}) is invariant under this reflection.\\

\begin{remark}[Logarithmic translations]
\label{rem:log-trans}
    Given a physical spacetime, the choice of unphysical spacetime in \cref{def:AH} is ambiguous upto a \(4\)-parameter family of \emph{logarithmic translations} at \(i^0\) which simultaneously change the \(C^{>1}\)-structure and the conformal factor at \(i^0\) \cite{Berg, Ash-log, Chr-log}. Let \(x^i = (t,x,y,z)\) be a \(C^{>1}\)-chart, at \(i^0\) such that \((M,g_{ab})\) is an unphysical spacetime satisfying \cref{def:AH} with conformal factor \(\Omega\). Let \(\rho^2 \defn \eta_{ij} x^i x^j\) with \(\eta_{ij}\) being the Minkowski metric at \(i^0\). Define another chart \(\tilde x^i\) at \(i^0\) by
\be\label{eq:log-trans}
    x^i = \tilde x^i (1 + 2 \ln\tilde\rho ~ \tilde x_j \lambda^j ) - \tilde\rho^2 \ln \tilde\rho~ \lambda^i
\ee
where \(\tilde\rho\) is defined in the new \(\tilde x^i\)-chart similar to \(\rho\), and \(\lambda^i\) are some constants. Consider the new unphysical manifold \(\tilde M\) which has a \(C^{>1}\)-structure in the \(\tilde x^i\)-chart, and the new unphysical metric \(\tilde g_{ab}\) and conformal factor \(\tilde\Omega\) given by
\be\label{eq:log-omega-factor}
    \tilde\Omega = \tilde\omega \Omega \eqsp \tilde g_{ab} = \tilde\omega^2 g_{ab} \quad\text{with~~} \tilde\omega \defn 1 - 2 \ln\rho~ x_i \lambda^i
\ee
 Then, it can be verified that \((\tilde M, \tilde g_{ab})\) with conformal factor \(\tilde\Omega\) also satisfies \cref{def:AH}. Note, that \(\partial x^i/\partial \tilde x^j\) is \(C^0\) at \(i^0\) but \(\partial^2 x^i/\partial \tilde x^j\partial \tilde x^k\) diverges logarithmically. Thus, \(M\) and \(\tilde M\) are equivalent as \(C^1\)-manifolds but \emph{not} as \(C^{>1}\)-manifolds at \(i^0\) (see \cref{def:C>1-struc}). Similarly, the relative conformal factor \(\tilde\omega\) in \cref{eq:log-omega-factor} does not satisfy the conditions in \cref{rem:freedom-Omega} since \(\tilde\omega\) is \(C^0\) at \(i^0\) but not \(C^{>0}\) (in either coordinate chart). Since \(M\) and \(\tilde M\) are equivalent \(C^1\)-manifolds, the tangent spaces at \(i^0\) can be naturally identified and the space of directions \(\hyp\) and \(\cyl\) are unaffected by these logarithmic translations. Thus, the logarithmic translations can also be parametrised by a \emph{direction-independent} vector \(\dd\lambda^a\) in \(Ti^0\) with components \(\lambda^i\) (in either coordinate chart).
\end{remark}

\section{Solutions for \(\dd E_{ab}\) on \(\cyl\)}
\label{sec:wave-hyp}

In this section we examine solutions to the Einstein equation \cref{eq:EE-hyp} on the hyperboloid \(\hyp\) and the cylinder \(\cyl\) at \(i^0\) using the construction in \cref{sec:dd}. 

The asymptotic unphysical Ricci tensor provides a scalar potential \(\dd E\) for \(\dd E_{ab}\) as follows (see \cite{AH} for details). Since, the metric \(g_{ab}\) is \(C^{>0}\), the tensor \(S_{ab}\) defined in \cref{eq:S-defn} is such that \(\Omh S_{ab}\) is \(C^{>-1}\) at \(i^0\) along spatial directions. Let \(\dd S_{ab}(\vec\eta) \defn \lim_{\to i^0} \Omh S_{ab}\), then,
\be\label{eq:E-potential}
    \dd E(\vec\eta) \defn \dd S_{ab}(\vec\eta) \dd\eta^a \dd\eta^b
\ee
defines a function \(\dd E\) intrinsic to \(\hyp\). \cref{eq:Weyl-S} in the limit to \(i^0\) along spatial directions, implies that \(\dd E\) acts as a scalar potential for \(\dd E_{ab}\) so that (see \cite{AH})
\be\label{eq:weyl-potential}
    \dd E_{ab} = -\tfrac{1}{4} \lb( \dd D_a \dd D_b \dd E + \dd E \dd h_{ab}  \rb)
\ee
Then, the second equation in \cref{eq:EE-hyp} is automatically satisfied and the first equation in \cref{eq:EE-hyp} gives
\be\label{eq:wave-hyp}
     \dd D^a \dd D_a \dd E + 3 \dd E = 0
\ee

We can solve \cref{eq:wave-hyp} using a decomposition \(\dd E = \sum\limits_{\ell,m} E_{\ell,m}(\alpha)Y_{\ell, m}(\theta^A)\) in terms of the spherical harmonic functions \(Y_{\ell, m}(\theta^A)\) so that each \(E_{\ell,m}(\alpha)\) satisfies (with the metric \cref{eq:h-hyp-coord})
\be\label{eq:red-wave-hyp}
    (1-\alpha^2)\frac{d^2}{d\alpha^2} E_{\ell,m} + \lb[ \ell(\ell+1) - \frac{3}{1-\alpha^2} \rb] E_{\ell,m} = 0
\ee
The solutions \(E_{\ell,m}(\alpha)\) are spanned by\footnote{We can also solve \cref{eq:red-wave-hyp} in terms of the \emph{Gau\ss\ hypergeometric functions}, \((1-\alpha^2)^{-\half} \times {}_2F_1(- \tfrac{\ell+2}{2},\tfrac{\ell-1}{2};\tfrac{1}{2};\alpha^2)\) and \(\alpha (1-\alpha^2)^{-\half} \times {}_2F_1(- \tfrac{\ell+1}{2},\tfrac{\ell}{2};\tfrac{3}{2};\alpha^2)\) which do not miss the solutions in \cref{eq:missed} but these obscure the parity transformation under \(\Upsilon\). These can be related to the solutions \cref{eq:Legendre-soln} using the transformation formulae in \S~3.2 \cite{special-func} or \S~14.3 \cite{DLMF}.}
\be\label{eq:Legendre-soln}
    (1-\alpha^2)^\half P^2_\ell (\alpha) \eqsp (1-\alpha^2)^\half Q^2_\ell(\alpha)
\ee
where \(P^2_\ell(\alpha)\) and \(Q^2_\ell(\alpha)\) are the \emph{Legendre functions} \cite{special-func}. Note, these miss out the solutions
\be\label{eq:missed}\begin{split}
    E_{\ell=0,m=0} & \propto (1+\alpha^2) (1-\alpha^2)^{-\half} \\
    E_{\ell=1,m=0,\pm 1} & \propto \alpha (3-\alpha^2) (1 - \alpha^2)^{-\half}
\end{split}\ee
Under the time reflection isometry \(\alpha \mapsto -\alpha\) on \(\hyp\), the solutions spanned by \(P^2_\ell(\alpha)\) and \(Q^2_\ell(\alpha)\) have a parity of \((-1)^{\ell}\) and \((-1)^{\ell+1}\), respectively. Combined with the parity \((-1)^\ell\) of the spherical harmonics \(Y_{\ell,m}\) under \(\theta^A \mapsto -\theta^A\), we get two linearly independent solutions \(\dd E^{\rsub{(odd)}}\) and \(\dd E^{\rsub{(even)}}\) to \cref{eq:wave-hyp} which are odd and even, respectively, under the reflection isometry \(\Upsilon: (\alpha, \theta^A) \mapsto (-\alpha, -\theta^A)\) of \(\hyp\). Thus the solutions to \cref{eq:wave-hyp} have the following behaviour in \(\alpha\)
\begin{subequations}\label{eq:wave-soln}\begin{align}
    \dd E^{\rsub{(odd)}} & = 
    \begin{cases}
    \text{constant} \times \alpha (1-\alpha^2)^{-\half} & \quad\quad\;\;\; \text{for } \ell = 0 \\
    E_1(\theta^A) (1 - \alpha^2)^{-\half} & \quad\quad\;\;\; \text{for } \ell = 1 \\
    E_2(\theta^A)(1-\alpha^2)^\half Q_\ell^2(\alpha) & \quad\quad\;\;\; \text{for } \ell \geq 2
    \end{cases} \\[2ex]
    \dd E^{\rsub{(even)}} & = 
    \begin{cases}
    \text{constant} \times (1+\alpha^2) (1-\alpha^2)^{-\half} & \text{for } \ell = 0 \\
    E_3(\theta^A) \alpha (3-\alpha^2) (1 - \alpha^2)^{-\half} & \text{for } \ell = 1 \\
     E_4(\theta^A) (1-\alpha^2)^\half P_\ell^2(\alpha) & \text{for } \ell \geq 2
    \end{cases}
\end{align}\end{subequations}
The solutions \(\dd E^{\rsub{(odd)}}\) for \(\ell = 0,1\) have \(\dd E_{ab} = 0\), and are ``pure-gauge'' solutions generated by logarithmic translations (see \cref{eq:E-transform-log-trans}). Computing the Spi-supermomenta \cref{eq:charge-hyp} where \(\dd f\) is a Spi-translation (see \cref{eq:trans-hyp-coord}), the solutions \(\dd E^{\rsub{(even)}}\) with \(\ell = 0,1\) represent the ``Schwarzschild-part'' of the solution which, in general, is boosted relative to the coordinates \((t,x,y,z)\) at \(i^0\).\\

Using the diffeomorphism between \(\hyp\) and \(\cyl \setminus \nulls^\pm\) we can treat \(\dd E\), and \(\dd E_{ab}\) as fields on \(\cyl\). For null-regular spacetimes we need \(\dd\Sigma^{-1} \dd E_{ab} \dd U^a \dd U^b\) to have a finite limit to \(\nulls^\pm\), where \(\alpha = \pm 1\). As noted above the reflection-odd solutions \(\dd E = \dd E^{\rsub{(odd)}}\) with \(\ell = 0,1\) do not contribute to \(\dd E_{ab}\), and an explicit computation, using \cref{eq:h-hyp-coord,eq:U-q-coord,eq:wave-soln}, shows that for the reflection-even solutions \(\dd E = \dd E^{\rsub{(even)}}\) 
\be\label{eq:E-cyl-falloff}
    \lim_{\to \nulls^\pm} \dd\Sigma^{-1} \dd E_{ab} \dd U^a \dd U^b \text{ exists} \eqsp
    \lim_{\to \nulls^\pm} \tilde{\dd q}_a{}^b \dd E_{bc} \dd U^c = \lim_{\to \nulls^\pm} \tilde{\dd q}_a{}^c \tilde{\dd q}_b{}^d \dd E_{cd} = 0
\ee
while for the reflection-odd solutions \(\dd E = \dd E^{\rsub{(odd)}}\) with \(\ell \geq 2\), these limits diverge. Thus for null-regular spacetimes \(\dd E_{ab}\) is reflection-even, and we have
\be\begin{split}
    \Upsilon \circ (\dd\Sigma^{-1} \dd E_{ab} \dd U^a \dd U^b)\vert_{\nulls^-} & = (\dd\Sigma^{-1} \dd E_{ab} \dd U^a \dd U^b)\vert_{\nulls^+}
\end{split}\ee 
in our choice of the rescaling function \cref{eq:Sigma-r,eq:Sigma-hyp}; for more general choices we get \cref{eq:reflection-electric}.\\

Consider the logarithmic translations as described in \cref{rem:log-trans} generated by a direction-independent vector \(\dd\lambda^a\) at \(i^0\). Any such vector can be written in the form (see for instance \cref{rem:trans-vectors})
\be\begin{aligned}
    \dd\lambda^a &= \dd\lambda \dd\eta^a + \dd D^a \dd\lambda \\
    \text{where } \dd\lambda & = \dd\lambda^a \dd\eta_a \text{ and } \dd D_a \dd D_b \dd\lambda + \dd h_{ab} \dd \lambda = 0
\end{aligned}\ee
Then, under the logarithmic translation generated by \(\dd\lambda^a\) we have \cite{Ash-log}
\be\label{eq:E-transform-log-trans}
    \dd E \mapsto \dd E + 4 \dd \lambda \eqsp \dd E_{ab} \mapsto \dd E_{ab}
\ee
Thus, the ``pure-gauge'' solutions \(\dd E^{\rsub{(odd)}}\) with \(\ell = 0,1\) in \cref{eq:wave-soln} are precisely the ones generated by a logarithmic translation.
 
Since the Spi-translations \(\dd f \in \mf t^0\) also satisfy \(\dd D_a \dd D_b \dd f + \dd f \dd h_{ab} = 0\) (\cref{eq:trans-hyp}) and can be represented by direction-independent vectors at \(i^0\) (\cref{rem:trans-vectors}), the Spi-translations \(\dd f\) on \(\hyp\) are also spanned by functions corresponding to \(\dd E^{\rsub{(odd)}}\) with \(\ell = 0,1\) in \cref{eq:wave-soln}, that is,\footnote{While both the logarithmic translations and Spi-translations are spanned by the same functions on \(\hyp\) and by direction-independent vectors at \(i^0\), their transformations on the Ashtekar-Hansen structure are very different. The logarithmic translations change the \(C^{>1}\)-structure at \(i^0\) (see \cref{rem:log-trans}) while Spi-translations preserve any chosen \(C^{>1}\)-structure. Similarly, the potential \(\dd E\) transforms as in \cref{eq:E-transform-log-trans} under logarithmic translations but is invariant under all Spi-supertranslations \cite{AH,Ash-log}.}
\be\label{eq:trans-hyp-coord}
    \dd f = (1-\alpha^2)^{-\half} \lb[ F_{0,0} \alpha + \sum_m F_{1, m} Y_{\ell=1,m} (\theta^A) \rb] \in \mf t^0
\ee
where \(F_{0,0}, F_{1,m}\), are constants. Using \cref{eq:trans-vector-defn} it can be checked these correspond to the direction-independent vectors \(\dd v^a\) at \(i^0\) spanned by constant linear combinations of the vectors \((\partial_t, \partial_x, \partial_y, \partial_z)\) determined by the \(C^{>1}\)-coordinates at \(i^0\). Note that \(\dd f\) diverges at \(\nulls^\pm\) (viewed as a function on \(\cyl\)) but the rescaled function \(\dd F = \dd\Sigma \dd f\) (with \(\dd\Sigma\) given by \cref{eq:Sigma-hyp}) is smooth at \(\nulls^\pm\) and
\be\label{eq:F-null-coord}
    \dd F\vert_{\nulls^\pm} = \pm F_{0,0} + \sum_m F_{1, m} Y_{\ell=1,m} (\theta^A)
\ee
Thus, the boundary values of \(\dd F\) on \(\nulls^\pm\) are precisely the \(\ell = 0,1\) functions on \(\bb S^2\) and are reflection-odd i.e. \(\Upsilon \circ \dd F\vert_{\nulls^-} = - \dd F\vert_{\nulls^+}\).

\section{Comparison to the Comp\`ere-Dehouck supermomenta on \(\hyp\)}
\label{sec:CD}

Using the symplectic formalism for general relativity, Comp\`ere and Dehouck \cite{CD} derived an expression for the supermomenta at \(i^0\), different from the one in \cref{eq:charge-hyp}. This expression was used by Troessaert to resolve the matching of supertranslations and supermomenta for linearised perturbations off Minkowski spacetime \cite{Tro}. In this section, we compare their results to the ones in the main body of the paper.

 The spacetimes considered in \cite{CD,Tro} are asymptotically-flat in the sense of \cref{def:AH} and satisfy the additional ``boundary condition'' \(\dd h^{ab} \dd K_{ab} = 0\),  where \(\dd K_{ab}\) is the tensor potential for the magnetic part of the asymptotic Weyl tensor on \(\hyp\) \cite{AH}. This condition restricts the Spi-supertranslations \(\mf s^0\) to the subalgebra given by \cite{CD}
\be\label{eq:st-hyp-cd}
    \mf s^0_\cd = \set{ \dd f \in \mf s^0 \st \dd D^a \dd D_a \dd f + 3 \dd f = 0 }
\ee
Following \cref{sec:wave-hyp}, we can obtain solutions for \(\dd f \in \mf s^0_\cd\) which again split into even and odd parity solutions \(\dd f^{\rsub{(even)}}\) and \(\dd f^{\rsub{(odd)}}\) as in \cref{eq:wave-soln}.

The supermomenta for any \(\dd f \in \mf s^0_\cd\) derived in \cite{CD} is then given by (in our notation)
\be\label{eq:charge-hyp-CD}
    \mc Q_\cd[\dd f; S] = \tfrac{1}{2} \int_S \dd\varepsilon_2  \lb( \dd E \dd D_a \dd f - \dd D_a \dd E \dd f  \rb) \dd u^a
\ee
where \(\dd E\) is the potential for \(\dd E_{ab}\) as defined in \cref{eq:E-potential}. Using \cref{eq:wave-hyp,eq:st-hyp-cd} it can be shown that the flux of the Comp\`ere-Dehouck supermomenta across any region \(\Delta\hyp\) bounded by two cross-sections vanishes, that is, the Comp\`ere-Dehouck supermomenta are exactly conserved on \(\hyp\).

The Comp\`ere-Dehouck supermomenta \cref{eq:charge-hyp-CD} are related to \cref{eq:charge-hyp} as follows. Using \cref{eq:wave-hyp}, for any \(\dd f \in \mf s^0_\cd\), we have
\be\begin{split}
    \dd E_{ab} \dd D^b \dd f & = \tfrac{1}{2} \lb( \dd E \dd D_a \dd f - \dd D_a \dd E \dd f  \rb) - \tfrac{1}{4} \dd D^b \dd E (\dd D_a \dd D_b \dd f + \dd h_{ab}\dd f) - \tfrac{1}{2} \dd D^b \lb( \dd D_{[a} \dd E \dd D_{b]} \dd f \rb) \\
\end{split}\ee
Integrating over a cross-section \(S\) of \(\hyp\), the last term vanishes as it is a boundary term on \(S \cong \bb S^2\) and we get
\be\label{eq:charge-hyp-compare}
    \mc Q[\dd f; S] = \mc Q_\cd [\dd f; S] - \tfrac{1}{4} \int_S \dd\varepsilon_2 ~ \dd D^b \dd E (\dd D_a \dd D_b \dd f + \dd h_{ab}\dd f) \dd u^a
\ee

Now, we consider the Comp\`ere-Dehouck supermomenta on \(\cyl\) and their limits to the space of null directions \(\nulls^\pm\). As before, we will consider null regular spacetimes where, using the explicit solutions in \cref{eq:wave-soln}, it can be shown that
\be\label{eq:E-potential-nulls}
    \lim_{\to \nulls^\pm} \dd\Sigma^{-1} \dd E_{ab} \dd U^a \dd U^b = \lim_{\to \nulls^\pm} - \frac{\partial^2}{~\partial\alpha^2} (\dd\Sigma \dd E)
\ee
exists whenever \(\dd E^{\rsub{(odd)}} = 0 \) for \(\ell \geq 2\), while the odd-solutions with \(\ell = 0,1\) do not contribute. In this case the only non-vanishing Comp\`ere-Dehouck supermomenta correspond to the odd-parity supertranslations \(\dd f = \dd f^{\rsub{(odd)}} \in \mf s^0_\cd\) \cite{CD,Tro}. Further, with \(\dd E = \dd E^{\rsub{(even)}}\) and \(\dd f = \dd f^{\rsub{(odd)}}\) it can be shown that
\be\label{eq:extra-falloff}
    \lim_{S \to \nulls^\pm} \int_S \dd\varepsilon_2 ~ \dd D^b \dd E (\dd D_a \dd D_b \dd f + \dd h_{ab}\dd f) \dd u^a = 0
\ee
where, for a Spi-translation \(\dd f\) this term identically vanishes (from \cref{eq:trans-hyp}), and for \(\dd f = \dd f^{\rsub{(odd)}}\) with \(\ell \geq 2\) the vanishing of the limit follows from the falloff of \(\dd E = \dd E^{\rsub{(even)}}\) in \cref{eq:wave-soln}. The limit of the Comp\`ere-Dehouck supermomenta to \(\nulls^\pm\) is then finite and given by \cite{Tro}
\be
    \mc Q_\cd [\dd f; \nulls^\pm] = - \int_{\nulls^\pm} \tilde{\dd\varepsilon}_2~ \dd F \frac{\partial^2}{~\partial\alpha^2} (\dd\Sigma \dd E)
\ee 
where, as before, \(\dd F = \dd\Sigma \dd f\). From \cref{eq:charge-null-C,eq:charge-hyp-compare,eq:E-potential-nulls,eq:extra-falloff} we see that, on \(\nulls^\pm\), the Comp\`ere-Dehouck supermomenta coincide with the Spi-supermomenta used the main body of the paper, that is,
\be
    \mc Q_\cd [\dd f; \nulls^\pm] = \mc Q [\dd f; \nulls^\pm]
\ee

Further, the odd-supertranslations represented by \(\dd F = \dd\Sigma \dd f^{\rsub{(odd)}}\) are fluxless on \(\cyl\) and satisfy \cref{eq:tot-fluxless-symm} and hence determine an equivalence class in the diagonal supertranslation algebra \(\mf s^\times\). Thus, the matching of the supertranslations and supermomenta can be, equivalently, derived using the Comp\`ere-Dehouck expression for the supermomenta. Thus, our result is a generalisation of the analysis by Troessaert \cite{Tro} to full nonlinear general relativity.

\section{Relation to some coordinate-based approaches}
\label{sec:BS}

In this appendix we collect the relations between our covariant approach to some of the coordinate-based approaches.

Given the Ashtekar-Hansen structure of \cref{def:AH}, consider a different choice of conformal factor \(\Omega_\bs = \varpi \Omega\) so that the Bondi condition \(\nabla_a^\bs \nabla_b^\bs \Omega_\bs\vert_{\scri^+} = 0\) holds. We denote quantities computed in this choice of conformal factor by a \(\bs\) for Bondi-Sachs. The Bondi-Sachs normal to \(\scri^+\) is then \(n^a_\bs \vert_{\scri^+} = \varpi^{-1} n^a\). The Bondi condition implies that \(\nabla_a^\bs n^a_\bs\vert_{\scri^+} = 0\) which in terms of the conformal-completion with \(\Omega\) gives \(\Lie_n \ln\varpi \vert_{\scri^+} = -\tfrac{1}{4} \nabla_a n^a \). From \cref{cond:Omega-at-i0} we have \(\nabla_a n^a\vert_{i^0} = 8\), and thus \(\varpi\vert_{\scri^+} = O(1/r)\), as we approach \(i^0\) along \(\scri^+\), where \(r\) is the \(C^{>0}\) ``radial'' coordinate at \(i^0\) from \cref{sec:dd}. The Bondi-Sachs parameter \(u\) on \(\scri^+\) is defined by \(n^a_\bs \nabla_a u = 1\) which gives \(u = O(1/r)\) with \(u \to - \infty\) being the limit to spatial infinity along \(\scri^+\). Note, however, that the unphysical metric in the Bondi-Sachs completion is \(g_{ab}^\bs = \varpi^2 g_{ab} = O( u^2) g_{ab}\) which diverges in any \(C^1\)-chart at \(i^0\) even in Minkowski spacetime.

From \cref{sec:dd}, any choice of the rescaling function behaves as \(\Sigma = O(1/r) = O(u)\) approaching \(i^0\) along \(\scri^+\). Converting to the conformal factor \(\Omega_\bs\), the null-regularity conditions of \cref{def:null-regular} and \cref{eq:S-falloff,eq:sigma-falloff} imply that
\be\label{eq:null-regular-BS}\begin{aligned}
    \lim_{u \to -\infty} \mc P^\bs  &\text{ is a smooth function on } \bb S^2 \\[1.5ex]
    N_{ab}^\bs = O (1/u^{1+\epsilon}) \eqsp \mc S_a^\bs &= O (1/u^{1+\epsilon}) \eqsp \mc R_{ab}^\bs = O(1/u^{2+\epsilon}) \\[1.5ex]
    \lim_{u \to -\infty} \sigma_{ab}^\bs &\text{ is bounded}
\end{aligned}\ee
for some small \(\epsilon > 0\). In the Newman-Penrose notation (see \cref{sec:non-Bondi}) as \(u \to -\infty\), \(\Re\psi_2^\bs\) has a limit as a smooth function on \(\bb S^2\), \(\psi_4^\bs\) fallsoff as \(O(1/u^{2+\epsilon})\), \(\psi_3^\bs\) and the News function \(N\) falloff as \(O(1/u^{1+\epsilon})\), and the spin-coefficient \(\sigma^\bs\) is bounded.

Consider now the vector field on \(\scri^+\), \(\xi^a = f_\bs n^a_\bs = f n^a\) where \(f_\bs\) is a smooth function on \(\bb S^2\) representing the BMS-supertranslation in a Bondi-Sachs frame. From the above discussion we have
\be\label{eq:st-BS-to-gen}
    f = \varpi^{-1} f_\bs = O(1/u) \implies \lim_{u \to -\infty} f = 0 
\ee 
The rescaled function \(F = \Sigma f\) behaves as \(F = O(u^0) f_{\bs}\) where \(O(u^0)\) denotes some smooth function on \(\bb S^2\). We can use the rescaling freedom in \(\Sigma\) (\cref{rem:freedom-Sigma}) so that \(\dd F^+ = \lim_{u \to -\infty } F = f_\bs\). Note that the remaining conformal freedom in \(f_\bs\) (conformal weight \(w=1\)) corresponds precisely to the rescaling freedom in \(\dd F^+\) (rescaling weight \(r=1\)). Thus, the BMS-supertranslations in terms of the limiting functions \(\dd F^+\) given in \cref{eq:st-null-lim} are precisely the usual BMS-supertranslation functions in the Bondi frame.

Similarly along spatial directions, we can relate the coordinates used in \cref{eq:g-i0} at \(i^0\) to the Beig-Schmidt radial coordinate \cite{Beig-Schmidt} by \(\rho_\bs = O(1/(r^2 - t^2))\) so that \(i^0\) lies at \(\rho_\bs \to \infty\) (now we use \(\bs\) for Beig-Schmidt). It was shown in \cite{Beig-Schmidt} that one can choose coordinates \((\rho_\bs, X^a)\) where \(X^a\) are some coordinates on the hyperboloids of \(\rho_\bs = \text{constant}\), so that the physical metric takes the form
\be
    \hat g_{ab} \equiv \lb[ 1 + \frac{\sigma^{(1)}}{\rho_\bs} \rb]^2 d\rho_\bs^2 + \rho_\bs^2 \lb[ h^{(0)}_{ab} + \frac{h^{(1)}_{ab}}{\rho_\bs} + o(1/\rho_\bs) \rb] d X^a d X^b + o(1/\rho_\bs)
\ee
The relation to the direction-dependent quantities on \(\hyp\) is given by \(\dd h_{ab} \equiv h^{(0)}_{ab}\), \(\dd E \equiv 4 \sigma^{(1)}\), while \(h^{(1)}_{ab}\) is related to the tensor potential for the magnetic part of the asymptotic Weyl tensor on \(\hyp\) (see Eqs.~3.30 and 3.31 \cite{Beig-Schmidt}).

In the Beig-Schmidt formalism Spi-supertranslations are given by the coordinate transformations (see Eqs.~2.3 and 2.4 \cite{Tro}; note that Eq.~2.18 \cite{Beig-Schmidt} seems to have a typographical error in the transformation of \(\rho_\bs\)) 
\be\label{eq:st-BS}\begin{aligned}
    \rho_\bs &\mapsto \rho_\bs \lb[ 1 - \tfrac{1}{\rho_\bs} f(X^a) + o(1/\rho_\bs) \rb] \\
    X^a &\mapsto X^a - \tfrac{1}{\rho_\bs} h^{(0)}{}^{ab} D_b f(X^a) + o(1/\rho_\bs)
\end{aligned}\ee
In terms of the \(C^{>1}\)-coordinates at \(i^0\) these correspond to the vector fields \cref{eq:st-vec-hyp} with \(\dd f \equiv f(X^a)\).

In Minkowski spacetime, the explicit transformations between the Bondi-Sachs and Beig-Schmidt coordinates reproduce the relation \cref{eq:st-null-regular} between the BMS-supertranslations and the Spi-supertranslations. For suitably regular linearised perturbations on Minkowski spacetime, these coordinate transformations also reproduce the relation \cref{eq:E-field-match} between the perturbed Weyl tensor components at spatial infinity. A similar approach was used in \cite{Tro} to prove the matching of supermomenta in the linearised theory on Minkowski spacetime using coordinates adapted to the formalism of Friedrich \cite{Friedrich}. Using the coordinate transformations detailed in \cite{Herb-Kerr} one can also consider linearised perturbations around a Kerr background though the computations are extremely tedious. Nevertheless such explicit coordinate transformations are not available in general spacetimes, and our covariant approach is more suited to the general matching problem.


\newpage

\bibliographystyle{JHEP}
\bibliography{BMS-matching}      
\end{document}